\documentclass[aps,preprint,showpacs]{revtex4}
\usepackage{graphicx}
\begin{document}
\newcommand{\be}{\begin{equation}}
\newcommand{\ee}{\end{equation}}
\newcommand{\bea}{\begin{eqnarray}}
\newcommand{\eea}{\end{eqnarray}}

\title{Fisher zeros and Potts zeros
of the $Q$-state Potts model for nonzero external magnetic field}
\author{Seung-Yeon Kim\footnote{Electronic address: sykim@kias.re.kr}}
\affiliation{School of Computational Sciences,
Korea Institute for Advanced Study,\\
207-43 Cheongryangri-dong, Dongdaemun-gu, Seoul 130-722, Korea}

%\date{\today}

\begin{abstract}
The properties of the partition function zeros in the complex temperature plane
(Fisher zeros) and in the complex $Q$ plane (Potts zeros)
are investigated for the $Q$-state Potts model in an arbitrary nonzero external
magnetic field $H_q$,
using the exact partition function of the one-dimensional model.
The Fisher zeros of the Potts model in an external magnetic field are discussed
for any real value of $Q\ge0$.
The Potts zeros in the complex $Q$ plane for nonzero magnetic field are studied
for the ferromagnetic and antiferromagnetic Potts models.
Some circle theorems exist for these zeros.
All Fisher zeros lie on a circle for $Q>1$ and $H_q\ge0$ except $Q=2$ (Ising model)
whose zeros lie on the imaginary axis.
All Fisher zeros also lie on a circle for any value of $Q$ when $H_q=0$
(except $Q=0$, 1 and 2) or $H_q=-\infty$ (except $Q=1$, 2 and 3).
All Potts zeros of the ferromagnetic model lie on a circle for $H_q\ge0$.
When $H_q=0$ or $-\infty$, all Potts zeros lie on a circle for both
the ferromagnetic and antiferromagnetic models.
All Potts zeros of the antiferromagnetic model with $H_q<0$ also lie on a circle
for $(x+1)^{-1}\le a<1$, where $a=e^{\beta J}$ and $x=e^{\beta H_q}$.
It is found that a part of the Fisher zeros or the Potts zeros lie on a circle
for the specific ranges of $H_q$.
It is shown that some Fisher or Potts zeros can cut the positive real axis.
Furthermore, the Fisher zeros or the Potts zeros lie on the positive real axis
for the specific ranges of $H_q$.
The densities of zeros are also calculated and discussed.
The density of zeros at the Fisher edge singularity diverges, and the edge
critical exponents at the singularity satisfy the scaling law.
There exists the Potts edge singularity in the complex $Q$ plane
which is similar to the Fisher edge singularity in the complex temperature plane.
\end{abstract}


\maketitle

%%%%%%%%%%%%%%%%%%%%%%%%%%%%%%%%%%%%%%%%%%%%%%%%%%%%%%%%%%%%%%%%%%%%%%%%%%%

\section{introduction}

The $Q$-state Potts model \cite{potts,wu82,wu84,baxter82a,martin91}
is a generalization of the Ising ($Q=2$) model.
The $Q$-state Potts model exhibits
a rich variety of critical behavior
and is very fertile ground for the analytical and numerical investigations
of first- and second-order phase transitions.
The Potts model is also related to other outstanding
problems in physics and mathematics.
Fortuin and Kasteleyn \cite{kasteleyn,fortuin} have shown that
the $Q$-state Potts model in the limit $Q\to1$ defines
the problem of bond percolation.
They \cite{fortuin} also showed that the problem of resistor network
is related to a $Q=0$ limit of the partition function of the Potts model.
In addition, the zero-state Potts model describes the statistics
of treelike percolation \cite{stephen}, and is equivalent to the undirected
Abelian sandpile model \cite{majumdar}.
The $Q={1\over2}$ state Potts model has a connection to a dilute spin glass
\cite{aharony}.
The $Q$-state Potts model with $0\le Q\le1$ describes
transitions in the gelation and vulcanization processes
of branched polymers \cite{lubensky}.

By introducing the concept of the zeros of the partition function
in the {\it complex} magnetic-field plane (Yang-Lee zeros),
Yang and Lee \cite{yang} proposed a mechanism
for the occurrence of phase transitions in the thermodynamic limit
and yielded a new insight into the unsolved problem of the ferromagnetic
Ising model in an arbitrary nonzero external magnetic field.
It has been known
\cite{yang,lee,itzykson,creswick97,creswick98,creswick99,janke01,janke02a,janke02b,kromhout}
that the distribution of the zeros of a model
determines its critical behavior.
Lee and Yang \cite{lee} also
formulated the celebrated circle theorem which states that
the partition function zeros of the Ising ferromagnet lie on the unit circle
in the complex magnetic-field ($x=e^{\beta H}$) plane.
However, for the ferromagnetic $Q$-state Potts model with $Q>2$
the Yang-Lee zeros in the complex $x$ plane
lie close to, but not on, the unit circle with the two
exceptions of the critical point $x=1$ ($H=0$) itself and the
zeros in the limit $T=0$ \cite{creswick99,kim98a,kim99,kim00a,kim02}.
It has been shown \cite{kim00a,kim02}
that the distributions of the ferromagnetic Yang-Lee zeros for $Q>1$
have similar properties independent of dimension.
Recently, the exact results on the Yang-Lee zeros of the ferromagnetic
Potts model have been found using the one-dimensional model
\cite{kim00a,mittag,glumac94}.
Mittag and Stephen \cite{mittag} studied the Yang-Lee zeros
of the three-state Potts ferromagnet in one dimension.
Glumac and Uzelac \cite{glumac94} found the eigenvalues of the transfer matrix
of the one-dimensional Potts model for general $Q$.
In particular, they have shown that the ferromagnetic Yang-Lee zeros can
lie on the positive real axis for $Q<1$.
Yang also calculated the Yang-Lee zeros of the one-dimensional Ising antiferromagnet,
and showed that they lie on the negative real axis \cite{yang2}.

The first zero, which we call the edge zero, and its complex conjugate of
a circular distribution of the Yang-Lee zeros of the Potts model
cut the positive real axis
at the physical critical point $x_c=1$ for $T\le T_c$ in the thermodynamic limit.
However, for $T>T_c$ the edge zero does not cut the positive real axis
in the thermodynamic limit, that is, there is a gap in the distribution of zeros
around the positive real axis. Within this gap, the free energy is analytic
and there is no phase transition.
Kortman and Griffiths \cite{kortman} carried out the first systematic investigation
of the magnetization at the edge zero, based on the high-field, high-temperature
series expansion for the Ising model on the square lattice and
the diamond lattice.
They found that above $T_c$ the magnetization at the edge zero
diverges for the square lattice and is singular for the diamond lattice.
For $T>T_c$ we rename the edge zero as the Yang-Lee edge singularity.
The divergence of the magnetization at the Yang-Lee edge singularity
means the divergence of the density of zeros, which does not occur
at a physical critical point.
Fisher \cite{fisher78} proposed the idea that the Yang-Lee edge singularity
can be thought of as a new second-order phase transition with associated critical
exponents and the Yang-Lee edge singularity can be considered
as a conventional critical point.
The critical point of the Yang-Lee edge singularity
is associated with a $\phi^3$ theory, different from the usual
critical point associated with the $\phi^4$ theory.
The crossover dimension of the Yang-Lee edge singularity is $d_c=6$.

Fisher \cite{fisher65} emphasized that the partition
function zeros in the complex temperature plane (Fisher zeros)
are also very useful in understanding phase transitions, and
showed that for the square lattice Ising model in the absence of
an external magnetic field the Fisher zeros in the complex $y=e^{-\beta J}$ plane
lie on two circles (the ferromagnetic circle $y_{\rm FM}=-1+\sqrt{2}e^{i\theta}$
and the antiferromagnetic circle $y_{\rm AFM}=1+\sqrt{2}e^{i\theta}$)
in the thermodynamic limit.
In particular, using the Fisher zeros both the ferromagnetic
phase and the antiferromagnetic phase can be considered
at the same time.
The critical behavior of the square-lattice Potts model in both
the ferromagnetic and antiferromagnetic phases
has been studied using the distribution of the Fisher zeros, and
the Baxter conjecture \cite{baxter82b} for the antiferromagnetic critical temperature
has been verified \cite{kim00b,kim01}.
Recently, the Fisher zeros of the $Q$-state Potts model on square
lattices have been studied extensively for integer $Q>2$ \cite
{martin91,kim01,janke01,janke02a,janke02b,maillard,martin83,martin85,martin86,wood,bhanot,alves91,chen,wu96,matveev96a,creswick98,kenna98a,kenna98b,kim98b,shrock00,chang01a,alves02,chang02a}
and noninteger $Q$ \cite{kim00b,kim01,shrock00,chang01a}.
Exact numerical studies have shown \cite
{martin91,kim00b,kim01,martin85,martin86,chen,matveev96a,creswick98,kim98b} that
for self-dual boundary conditions the Fisher zeros of the $Q>1$ Potts models
on a finite square lattice are located on the unit circle in the complex
$p$ plane for ${\rm Re}(p)>0$, where $p=(e^{\beta J}-1)/\sqrt{Q}$.
It has been analytically shown that all the Fisher zeros of the
infinite-state Potts model lie on the unit circle for any size of
square lattice with self-dual boundary conditions \cite{wu96}, and
the Fisher zeros near the ferromagnetic critical point of the
$Q>4$ Potts models on the square lattice lie on the unit circle
in the thermodynamic limit \cite{kenna98a,kenna98b}.
Chen {\it et al.} \cite{chen} conjectured that when $Q$ reaches a certain
critical value $\tilde{Q}_c(L)$, all Fisher zeros for $L\times L$ square
lattices with self-dual boundary conditions are located at the unit
circle $|p|=1$. Later, Kim and Creswick \cite{kim01} verified this conjecture
and found that $\tilde{Q}_c(L)$ approaches infinity in the thermodynamic limit.
The thermal exponent $y_t$ of the
square-lattice Potts antiferromagnet has also been studied
using the Fisher zeros near the antiferromagnetic critical point \cite{kim00b,kim01}.

The Yang-Lee zeros in the complex $x$ plane for real $y$ have been
extensively studied and well understood. However, the Fisher zeros
in the complex $y$ plane for real $x$ are much less well understood than the
Yang-Lee zeros in the complex $x$ plane.
The Fisher zeros in an external magnetic field have been considered
by Itzykson {\it et al.} \cite{itzykson} for the first time.
They studied the movement of the Fisher zero closest to the positive
real axis for the Ising model as the strength of a magnetic field ($x>1$)
changes. For nonzero magnetic field there is a gap in the distribution
of the Fisher zeros of the Ising model around the positive real axis
even in the thermodynamic limit, which means that there is no
phase transition.
Matveev and Shrock \cite{matveev95,matveev96b} studied the Fisher zeros of the
two-dimensional Ising model in an external magnetic field ($x>1$)
using the high-field, low-temperature series expansion
and the partition functions of finite-size systems.
They observed that the thermodynamic
functions and the density of zeros diverge at the edge zero when $x>1$,
and that the values of the critical exponents associated with the edge zero
for $x>1$ are nearly independent of $x$ and satisfy the Rushbrooke scaling law
approximately. We call the edge zero in the complex temperature plane
for $x>1$ as the Fisher edge singularity.
Kim and Creswick\cite{kim98b} extended the study of the
Fisher edge singularity to the $Q$-state Potts model ($Q\ge3$) on the square lattice.
They found that the values of the edge critical exponents at the Fisher edge singularities
are nearly independent of $Q$ ($Q\ge2$).
The critical behavior at the Fisher edge singularity
has also been studied for the Ising model and the three-state Potts model
on the simple-cubic lattice using the high-field, low-temperature
series expansions \cite{kim02}.
However, the Fisher zeros of the $Q$-state Potts model in an external magnetic
field have never been investigated for general $Q$ except few integer values of $Q$.

The partition function of the $Q$-state Potts model in the absence of an external
magnetic field is also known as the Tutte dichromatic polynomial \cite{tutte}
or the Whitney rank function \cite{whitney}
in graph theory and combinatorics of mathematics.
One of the most interesting properties of the antiferromagnetic Potts model is
that for $Q>2$ the ground-state is highly degenerate and
the ground-state entropy is nonzero.
The ground states of the antiferromagnetic Potts model are
equivalent to the chromatic polynomials \cite{birkhoff} in mathematics,
which play a central role in the famous four-color problem \cite{appel}.
The partition function zeros in the complex $Q$ plane (Potts zeros) of the
$Q$-state Potts model have been studied both in mathematics and in physics.
The Potts zeros at $\beta J=-\infty$, corresponding to $T=0$
for the antiferromagnetic model, have
been investigated extensively to understand the chromatic polynomials and
the ground states of the antiferromagnetic Potts model
\cite{kim01,chang02a,hall,berman,biggs,beraha79,beraha80,farrell,baxter86,baxter87,wakelin,brenti,shrock97a,shrock97b,shrock97c,shrock97d,shrock98a,shrock98b,rocek98a,rocek98b,tsai,brown98a,brown98b,shrock99a,shrock99b,sokal,chang01b,chang02b,salas,jacobsen}.
Recently, the Potts zeros at finite temperatures have been studied for
ladder graphs \cite{shrock00,chang02a}, $L\times L$ square lattices \cite{kim01},
and the model with long-range interactions \cite{glumac02}.
However, the properties of the Potts zeros for nonzero magnetic field have
never been known until now.

In this paper, we investigate the exact results on the partition function zeros
of the ferromagnetic and antiferromagnetc $Q$-state Potts models in one dimension,
and we unveil the unknown properties of the Fisher zeros and
the Potts zeros in an external magnetic field.
In the next section we obtain two master equations to determine
the partition function zeros of the one-dimensional Potts model.
In Sec. III we calculate and discuss the Fisher zeros of the $Q$-state Potts model
for any value of $Q$ and in an arbitrary magnetic field.
In Sec. IV we study the Potts zeros in the complex $Q$ plane of the ferromagnetic
and antiferromagnetic Potts models in an arbitrary external magnetic field.

%%%%%%%%%%%%%%%%%%%%%%%%%%%%%%%%%%%%%%%%%%%%%%%%%%%%%%%%%%%%%%%%%%%%%%%%%%%%%%%%%

\section{The $Q$-state Potts model}

The $Q$-state Potts model in an external magnetic field $H_q$
on a lattice $G$ with $N_s$ sites and $N_b$ bonds
is defined by the Hamiltonian
\be
{\cal H}_Q=-J\sum_{\langle i,j\rangle}\delta(\sigma_i,\sigma_j)
-H_q\sum_k\delta(\sigma_k,q),
\ee
where $J$ is the coupling constant (ferromagnetic model for $J>0$
and antiferromagnetic model for $J<0$),
$\langle i,j\rangle$ indicates a sum over nearest-neighbor pairs,
$\sigma_i=1,2,...,Q$,
and $q$ is a fixed integer between 1 and $Q$.
The partition function of the model is
\be
Z_Q=\sum_{\{ \sigma_n \}} e^{-\beta{\cal H}_Q},
\ee
where $\{\sigma_n\}$ denotes a sum over $Q^{N_s}$ possible spin
configurations and $\beta=(k_B T)^{-1}$.
The partition function can be written as
\be
Z(a,x,Q)=\sum_{E=0}^{N_b}\sum_{M=0}^{N_s}\Omega_Q(E,M) a^E x^M,
\ee
where $a=y^{-1}=e^{\beta J}$, $x=e^{\beta H_q}$,
$E$ and $M$ are positive integers $0\le E\le N_b$ and $0\le M\le N_s$,
respectively, and $\Omega_Q(E,M)$ is the number of states
with fixed $E$ and fixed $M$.
The states with $E=0$ ($E=N_b$) correspond to the antiferromagnetic
(ferromagnetic) ground states.
The parameter $Q$ enters the Potts model as an integer.
However, the study of the $Q$-state Potts model has been extended to
continuous $Q$ due to the Fortuin-Kasteleyn representation
of the partition function \cite{kasteleyn,fortuin}
and its extension \cite{blote}.

For the one-dimensional Potts model in an external field the
eigenvalues of the transfer matrix were found by Glumac and
Uzelac \cite{glumac94}. The eigenvalues are $\lambda_\pm=(A\pm i B)/2$,
where $A=a(1+x)+Q-2$ and $B=-i\sqrt{[a(1-x)+Q-2]^2+4(Q-1)x}$,
and $\lambda_0=a-1$ which is $(Q-2)$-fold degenerate.
The partition function of the one-dimensional model ($N=N_s=N_b$) is given by
\be
Z_N(a,x,Q)=\lambda_+^N+\lambda_-^N+(Q-2)\lambda_0^N.
\ee

When $\lambda_+$ and $\lambda_-$ are two dominant eigenvalues, we have
\be
Z_N\simeq\lambda_+^N+\lambda_-^N
\ee
for large $N$.
If we define $A=2C\cos\psi$ and $B=2C\sin\psi$,
where $C=\sqrt{(a-1)(a+Q-1)x}$, then $\lambda_\pm=C\exp(\pm i\psi)$,
and the partition function is
\be
Z_N=2C^N \cos N\psi.
\ee
The zeros of the partition function are then given by
\be
\psi=\psi_k={2k+1\over2N}\pi,\ \ \ \ \ k=0,1,2,...,N-1.
\ee
In the thermodynamic limit the locus of the partition function zeros is determined
by the solution of
\be
A=2C\cos\psi,
\ee
where $0\le\psi\le\pi$.
In the special case $Q=2$ the contribution by the eigenvalue $\lambda_0$
disappears from the partition function, Eq.~(4),
and the equation (8) determines all the locus even for finite systems.

On the other hand, when $\lambda_+$ and $\lambda_0$ are two dominant eigenvalues,
we have
\be
Z_N\simeq\lambda_+^N+(Q-2)\lambda_0^N
\ee
for large $N$. The partition function zeros are then determined by
\be
{\lambda_+\over\lambda_0}=(2-Q)^{1/N}\exp(i\phi),
\ee
where
\be
\phi=\phi_k={2\pi k\over N},\ \ \ \ \ k=0,1,2,...,N-1.
\ee
In the thermodynamic limit the locus of the partition function zeros is determined
by the solution of
\be
a^2x^2+(Q-1)x-a x A+(a-1)A e^{i\phi}-(a-1)^2e^{2i\phi}=0,
\ee
where $0\le\phi\le2\pi$.
The equation (12) also determines the locus of the zeros
when $\lambda_-$ and $\lambda_0$ are two dominant eigenvalues.

%%%%%%%%%%%%%%%%%%%%%%%%%%%%%%%%%%%%%%%%%%%%%%%%%%%%%%%%%%%%%%%%%%%%%%%%%%%%%%%%%

\section{Fisher zeros}

From the equation (8) the locus of the Fisher zeros is obtained to be
\be
y_1(\psi)={(Q-2)(x\cos2\psi-1)\pm i2
\sqrt{f}\over(Q-2)^2+4(Q-1)x\cos^2\psi},
\ee
where $f=x\cos^2\psi[(Q+x-1)(Q x-x+1)-Q^2x\cos^2\psi]$.
The edge zeros of $y_1(\psi)$ are given by
\be
y_\pm=y_1(0)={(Q-2)(x-1)\pm i2|x-1|\sqrt{(Q-1)x}\over(Q-2)^2+4(Q-1)x}.
\ee
In the absence of an external field, $H_q=0$ ($x=1$),
these edge zeros cut the real axis at the origin,
\be
y_\pm=0,
\ee
which corresponds to the $T=0$ ferromagnetic transition point.
If $f<0$, the zeros of $y_1(\psi)$ lie on the real axis.
However, if $f>0$, the zeros of $y_1(\psi)$ lie on a circle
\be
y_1(\psi)=y_c+D e^{\pm i\theta(\psi)}
\ee
in the complex $y$ plane, where $y_c$ (the center of the circle) and
$D$ are given by
\be
y_c={1\over2(Q-2)}{(Q-2)^2(x-1)^2-E(x+1)^2\over(Q-2)^2(x-1)+E(x+1)}
\ee
and
\be
D={1\over2(Q-2)}{(Q-2)^2(3x+1)(x-1)+E(x+1)^2
\over(Q-2)^2(x-1)+E(x+1)}.
\ee
$E$ is defined by
\be
E=(Q-2)^2+4(Q-1)x,
\ee
the argument $\theta$ is given by
\be
\cos\theta(\psi)={1\over D}
\bigg[{(Q-2)(x\cos2\psi-1)\over(Q-2)^2+4(Q-1)x\cos^2\psi}-y_c\bigg],
\ee
and the radius $r$ of the circle is
\be
r=|D|.
\ee
The one point of the circle $y_1(\psi)$,
\be
y_1\bigg(\psi={\pi\over2}\ {\rm or}\ \theta=\pi\bigg)={x+1\over2-Q},
\ee
always lies on the real axis.
The point $y_1({\pi\over2})$ lie on the positive real axis
for $Q<2$ and on the negative real axis for $Q>2$.

On the other hand, the equation (12) gives the second locus of the Fisher zeros
\be
y_2(\phi)={e^{i2\phi}-(x+1)e^{i\phi}+x\over e^{i2\phi}+(Q-2)e^{i\phi}-(Q-1)x}.
\ee
The two points of $y_2(\phi)$,
\be
y_2(0)=0
\ee
and
\be
y_2(\pi)={2(x+1)\over(3-Q)+(1-Q)x},
\ee
can be lie on the real axis.
From $|\lambda_\pm|=|\lambda_0|$ we also obtain
\be
y_\ast={1-x\over(Q-1)x+1}.
\ee

%%%%%%%%%%%%%%%%%%%%%%%%%%%%%%%%%%%%%%%%%%%%%%%%%%%%%%%%%%%%%%%%%%%%%%%%%%%%%%%%%%%

\subsection{$Q>1$}

In the special case $Q=2$ all the Fisher zeros lie on the imaginary axis,
and they meet the real axis at the origin only for $x=1$.
At $x=1$ ($H_q=0$), the loci $y_1(\psi)$ and $y_2(\phi)$ become
the identical locus as a circle \cite{shrock00}
\be
y(\rho)=-{1\over Q-2}+{1\over Q-2}e^{i\rho}
\ee
for any value of $Q$ except $Q=0$, 1, and 2.
On this circle three eigenvalues have the same magnitude
\be
|\lambda_+|^2=|\lambda_-|^2=|\lambda_0|^2=
{(Q-1)(Q-1-2\cos\rho)+1\over2(1-\cos\rho)}.
\ee
This circle cuts the real axis at two points
\be
y(\rho=0)=0
\ee
and
\be
y(\rho=\pi)={2\over 2-Q}.
\ee
The point $y(\rho=0)$ is the $T=0$ ferromagnetic transition point.
The point $y(\rho=\pi)$ has no physical meaning for $Q>2$,
but it may correspond to an antiferromagnetic transition point for $Q<2$
because the physical interval is
\be
1\le y\le\infty \ \ (\infty\ge T\ge0)
\ee
for antiferromagnetic interaction $J<0$.
For $x>1$ ($H_q>0$) all the Fisher zeros lie on the circle $y_1(\psi)$
($f>0$ for $Q>1$ and $x>1$) ,
whereas for $x<1$ ($H_q<0$) they are located on the loop $y_2(\phi)$.

Figure 1 shows the locus of the Fisher zeros in the complex $y$ plane
of the three-state Potts model for $x=1$, 2 and 3.
For $Q=3$ the center $y_c$ and the radius $r$ of the circle $y_1(\psi)$
are given by
\be
y_c=-1,\ \ r=1\ \ (x=1),
\ee
\be
y_c=-{19\over13},\ \ r={20\over13}\ \ (x=2),
\ee
and
\be
y_c=-{33\over17},\ \ r={35\over17}\ \ (x=3).
\ee
For $x=1$ two edge zeros $y_\pm$ cut the real axis at the origin
which is the $T=0$ ferromagnetic transition point.
However, as $x$ increases, they move away from the origin,
and there is a gap in the distribution of zeros, centered at $\theta=0$,
that is, the density of zeros is zero, $g(\theta)=0$, for $|\theta|<\theta_0$.
The edge angle $\theta_0$ ($=\theta(\psi=0)$) is given by
\be
\cos\theta_0={1\over D}
\bigg[{(Q-2)(x-1)\over E}-y_c\bigg],
\ee
and the edge zeros are expressed as
\be
y_\pm=y_c+De^{\pm i\theta_0}.
\ee
For example, for the three-state Potts model, the edge zeros are located at
\be
y_\pm=0 \ \ (x=1),
\ee
\be
y_\pm={1\over17}(1\pm4i) \ \ (x=2),
\ee
and
\be
y_\pm={1\over25}(2\pm4\sqrt{6}i) \ \ (x=3).
\ee

On the circle $y_1(\psi)$, the density of zeros $g(\theta(\psi))$
is given by
\be
g(\theta)={V|\sin\theta|\over2\pi\sqrt{(Q-2)(x-1)-E(y_c+D\cos\theta)}},
\ee
where $V$ is defined by
\be
V={D[(Q-2)^3+2(Q-1)(Q-2)(x+1)]\over[(Q-2)-2(Q-1)(y_c+D\cos\theta)]
\sqrt{(Q-2)(x+1)+(Q-2)^2(y_c+D\cos\theta)}}.
\ee
For $x=1$ the density of zeros reduces to a simple form
\be
g(\theta)={1\over2\pi}{Q|Q-2|\over(Q-2)^2+2(Q-1)(1-\cos\theta)},
\ee
which gives
\be
g(0)={1\over2\pi}{Q\over|Q-2|}
\ee
at the edge zeros $y_\pm=0$, showing a first-order transition.

On the other hand, for $x>1$ the density of zeros near the edge zeros
is expressed as
\be
g(\theta\sim\theta_0)\sim{1\over\sqrt{y(\theta)-y_\pm(\theta_0)}},
\ee
that is, the density of zeros diverges at the Fisher edge zeros
$y_\pm$ for $x>1$.
In this case, the Fisher edge zero is called the Fisher edge
{\it singularity} because of the divergence of the density of zeros.
The {\it edge} critical exponents $\alpha_e$, $\beta_e$ and $\gamma_e$
associated with the Fisher edge singularity are defined in the usual way,
\be
C_e\sim(y-y_\pm)^{-\alpha_e},
\ee
\be
m_e\sim(y-y_\pm)^{\beta_e},
\ee
and
\be
\chi_e\sim(y-y_\pm)^{-\gamma_e},
\ee
where $C_e$, $m_e$, and $\chi_e$ are the singular parts of
the specific heat, magnetization, and susceptibility, respectively.
The density of zeros near the Fisher edge singularity
is also given by
\be
g(\theta\sim\theta_0)\sim(y-y_\pm)^{1-\alpha_e}.
\ee
The Fisher edge singularity is characterized
by the edge critical exponents $\alpha_e={3\over2}$, $\beta_e=-{1\over2}$,
and $\gamma_e={3\over2}$. These values satisfy the Rushbrooke scaling law
\cite{kim02,kim98b}
\be
\alpha_e+2\beta_e+\gamma_e=2.
\ee

Figure 2 shows the locus of the Fisher zeros in the complex $y$ plane
of the $Q={3\over2}$ state Potts model for $x=1$ and $x=3$.
For $Q={3\over2}$ the values of the center $y_c$, the radius $r$,
and the edge zeros $y_\pm$ of the circle $y_1(\psi)$ are
\be
y_c=2,\ \ r=2,\ \ y_\pm=0 \ \ (x=1)
\ee
and
\be
y_c={66\over17},\ \ r={70\over17},\ \ y_\pm={4\over25}(-1\pm2\sqrt{6}i)\ \ (x=3).
\ee
The overall behavior in the figure 2 is similar to that in the figure 1.
However, there is a big difference in the location of the point
$y_1$($\psi={\pi\over2}$ or $\theta=\pi$), Eq.~(22), of the circle
$y_1(\theta(\psi))$.
The point $y_1(\pi)$ lies on the negative real axis for $Q>2$,
whereas it lies on the negative real axis for $1<Q<2$.
For $1<Q<2$ and $x\ge1$ we have
\be
y_1(\pi)>2,
\ee
and the point $y_1(\pi)$ may correspond to an antiferromagnetic transition point.

In the limit $H_q\to\infty$ ($x\to\infty$) the positive field
$H_q$ favors the state $q$ for every site and the $Q$-state Potts
model is transformed into the one-state model.
The Fisher zeros are determined by
\be
Z(y,x\to\infty,Q)\sim\sum_{E=0}^{N_b}\Omega_Q(E,M=N_s)y^{-E}=0
\ee
for any $Q$.
Because $\Omega_Q(E,M=N_s)=1$ for $E=N_b$ and 0 otherwise, Eq.~(53) is
\be
y^{-N_b}=0.
\ee
As $x$ increases, $|y|$ for all the zeros increases without bound \cite{kim98b}.

Now we turn to the distributions of Fisher zeros for $x<1$ ($H_q<0$)
which are completely determined by the loop $y_2(\phi)$.
The locus cuts the real axis at two points $y_2(0)$ and $y_2(\pi)$.
The point
\be
y_2(0)=0
\ee
is the $T=0$ ferromagnetic transition point.
For $Q\ge3$ the point $y_2(\pi)$ lies on the negative real axis,
and it has no physical meaning.
Figure 3 shows the locus of the Fisher zeros in the complex $y$ plane
of the three-state Potts model for $x={1\over2}$,
which gives
\be
y_2(\pi)=-3.
\ee
For $1<Q<2$, the point $y_2(\pi)$ lies on the positive real axis,
and it may be considered an antiferromagnetic transition point
because
\be
y_2(\pi)>1.
\ee
Figure 4 shows the locus of the Fisher zeros in the complex $y$ plane
of the $Q={3\over2}$ state Potts model for $x={1\over2}$.
For $Q={3\over2}$ and $x={1\over2}$ we obtain
\be
y_2(\pi)={12\over5}.
\ee
For $2<Q<3$, the value of $y_2(\pi)$ diverges at
\be
\bar{x}_1={3-Q\over Q-1}.
\ee
For $2<Q<3$ and $\bar{x}_1<x<1$, the point $y_2(\pi)$ lies
on the negative real axis, and the distributions of the Fisher zeros
show similar behavior to that in Fig.~3.
However, for $2<Q<3$ and $x<\bar{x}$, $y_2(\pi)>1$,
and the distributions of the zeros are similar to that in Fig.~4.

In the limit $H_q\to-\infty$ ($x\to0$) the partition function
of the $Q$-state Potts model, Eq.~(4), becomes
\be
Z(a,x=0,Q)=(a+Q-2)^N+(Q-2)(a-1)^N.
\ee
For an external field $H_q<0$, one of the Potts states is supressed
relative to the others, and the symmetry of the Hamiltonian is that of the
$(Q-1)$-state Potts model in zero external field.
Therefore, the partition function $Z(a,x=0,Q+1)$ of the ($Q+1$) state Potts model
in the limit $H_q\to-\infty$ is the same as the partition function
\be
Z(a,x=1,Q)=(a+Q-1)^N+(Q-1)(a-1)^N.
\ee
of the $Q$-state Potts model in the absence of an external magnetic field.
As $x$ decreases from 1 to 0, the $Q$-state Potts model is transformed
into the ($Q-1$)-state Potts model in zero external field \cite{kim98b}.
At $x=0$, the Fisher zeros of the $Q$-state Potts model lie on a circle
\be
y(\rho)=-{1\over Q-3}+{1\over Q-3}e^{i\rho}
\ee
for any value of $Q$ except $Q=1$, 2, and 3.
The zeros lie on the imaginary axis for the three-state Potts model at $x=0$.

%%%%%%%%%%%%%%%%%%%%%%%%%%%%%%%%%%%%%%%%%%%%%%%%%%%%%%%%%%%%%%%%%%%%%%%%%%%%%%%%%

\subsection{$Q<1$ and $x>1$}

For $x<\bar{x}_2$, where
\be
\bar{x}_2={1\over1-Q},
\ee
the locus consists of the loop $y_2(\phi)$ ($\phi_\ast\le\phi\le2\pi-\phi_\ast$) and
the line $y_1(\psi)$ ($0\le\psi\le\psi_\ast$) inside the loop $y_2$.
The line $y_1(\psi)$ lies on the real axis
between $y_\ast=y_1(\psi_\ast)$ and $y_-=y_1(0)$ ($y_\ast<y_-<0$),
and meets with the loop $y_2(\phi)$ at the point
\be
y_\ast=y_2(\phi_\ast)=y_2(2\pi-\phi_\ast),
\ee
where $|\lambda_+|=|\lambda_-|=|\lambda_0|$.
The loop $y_2(\phi)$ also cuts the real axis at the point $y_2(\pi)$ ($>1$).
Figure 5 shows the locus of the Fisher zeros for $Q={1\over2}$ and $x={3\over2}$
which has
\be
y_\ast=-2,\ y_-=1-{2\over\sqrt{3}}=-0.155,\ y_2(\pi)={20\over13}.
\ee

At $x=\bar{x}_2$, $y_\ast=-\infty$, and the line $y_1(\psi)$ lies on the
real axis between $-\infty$ and $y_-$ ($<0$).
In the region $\bar{x}_2<x<\bar{x}_3$, where
\be
\bar{x}_3={1+\sqrt{Q}\over1-\sqrt{Q}},
\ee
the line $y_1(\psi)$ lies outside the loop $y_2(\phi)$.
The line $y_1$ lies on the real axis between $-\infty$ and $y_-$ ($<0$)
and between $y_\ast$ ($>1$) and $\infty$.
The loop $y_2(\phi)$ cuts the real axis at the point $y_2(\pi)$,
and meets with the line $y_1(\psi)$ at $y_\ast$ ($1<y_2(\pi)<y_\ast$).
Figure 6 shows the locus of the Fisher zeros for $Q={1\over2}$ and $x=3$,
from which we obtain
\be
y_-={4\over15}(3-2\sqrt{6})=-0.506,\ y_2(\pi)=2,\ y_\ast=4.
\ee

At $x=\bar{x}_3$, $\phi_\ast=\pi$,
the other edge zero $y_+$ appears, and the loop $y_2(\phi)$ shrinks
to the point
\be
y_+=y_\ast=y_2(\pi)={-2\sqrt{Q}\over Q+(Q-2)\sqrt{Q}}.
\ee
For $x>\bar{x}_3$, the only locus is the line $y_1(\psi)$
on the real axis between $-\infty$ and $y_-$ ($<0$)
and between $y_+$ ($>1$) and $\infty$.
At $Q=0$, two edge zeros are simply expressed as
\be
y_\pm={1\pm\sqrt{x}\over2}
\ee
for $x>1$, and all the Fisher zeros still lie on the real axis.

%%%%%%%%%%%%%%%%%%%%%%%%%%%%%%%%%%%%%%%%%%%%%%%%%%%%%%%%%%%%%%%%%%%%%%%%%%%%%%%%%

\subsection{$Q<1$ and $x<1$}

For $x>1-Q$, the locus consists of
the loop $y_2(\phi)$ ($-\phi_\ast\le\phi\le\phi_\ast$),
the line $y_1(\psi)$ ($0\le\psi\le\psi_0$ and $\pi-\psi_0\le\psi\le\psi_\ast$)
on the real axis between $y_\ast=y_1(\psi_\ast)$ and $y_-=y_1(0)$ ($0<y_\ast<y_-<1$),
and the circle $y_1(\psi)$ ($\psi_0\le\psi\le\pi-\psi_0$).
The loop $y_2$ meets with the line $y_1$ at the point
\be
y_\ast=y_2(\phi_\ast)=y_2(-\phi_\ast),
\ee
where $|\lambda_+|=|\lambda_-|=|\lambda_0|$.
The circle $y_1(\psi)$ cuts the real axis at two points $y_1^0$ and
$y_1\big({\pi\over2}\big)$ ($y_\ast<y_1^0<y_-<y_1\big({\pi\over2}\big)$),
where the points $y_1^0$ and $y_1\big({\pi\over2}\big)$ are given by
\be
y_1^0=y_1(\psi_0)=y_1(\pi-\psi_0)=y_c+D
={2(Q-2)x(x-1)\over(Q-2)^2(x-1)+E(x+1)}
\ee
and
\be
y_1\bigg({\pi\over2}\bigg)=y_c-D={x+1\over2-Q}>1,
\ee
respectively. Similarly, the loop also cuts the real axis at two points
$y_2(0)$ and $y_\ast$ ($0=y_2(0)<y_\ast$).
For example, figure 7 shows the locus for $Q={1\over2}$ and $x={7\over10}$.
For these values we obtain
\be
y_\ast={6\over13},\ y_1^0={9\over11},\
y_-={1\over17}\Bigg(9+6\sqrt{7\over5}\Bigg)=0.947,\
y_1\bigg({\pi\over2}\bigg)={17\over15},
\ee
and the center $y_c$ and the radius $r$ for the circle
\be
y_c={161\over165},\ r={26\over165}.
\ee

At $x=1-Q$, the circle $y_1(\psi)$ shrinks to the edge zero $y_-=1$.
In the region $\bar{x}_4<x\le1-Q$,
where
\be
\bar{x}_4={1-\sqrt{Q}\over1+\sqrt{Q}},
\ee
the locus consists of the loop $y_2(\phi)$ and the line $y_1(\psi)$ on the
real axis between $y_\ast$ and $y_-$ ($y_2(0)<y_\ast<y_-\le1$).
The loop $y_2$ meets with the line $y_1$ at the point $y_\ast$,
and also cuts the real axis at the point $y_2(0)$ ($=0$).
At $x=\bar{x}_4$, the line $Q_1(\phi)$ shrinks to the point
\be
y_-=y_\ast=y_2(\pi)={2\sqrt{Q}\over Q+(2-Q)\sqrt{Q}}.
\ee
For $x<\bar{x}_4$,
the line disappears, and the only locus is the loop $y_2(\phi)$
which cuts the real axis at two points $y_2(0)$ and $y_2(\pi)$
($0=y_2(0)<y_2(\pi)<1$).
At $Q=0$, all the Fisher zeros lie only on the loop $y_2(\phi)$ for $x<1$.

%%%%%%%%%%%%%%%%%%%%%%%%%%%%%%%%%%%%%%%%%%%%%%%%%%%%%%%%%%%%%%%%%%%%%%%%%%%%%%

\subsection{One-state (Q=1) Potts model}

At $Q=1$, the partition function becomes
\be
Z_N(a,x,Q=1)=(ax)^N.
\ee
Therefore, all the Fisher zeros lie on the point $a=y^{-1}=0$.
On the other hand, in the limit $Q\to1$,
the loop $y_2(\phi)$ is reduced to
\be
y_2(\phi)=1-x e^{-i\phi},
\ee
and the Fisher zeros are uniformly distributed on this circle.
The circle $y_2(\phi)$ cuts the real axis at two points $1+x$ and $1-x$.
The point $1+x$ may be an antiferromagnetic transition point,
whereas the point $1-x$ can correspond to a ferromagnetic transition point for $x<1$.
The locus $y_1(\psi)$ also becomes the identical circle
with the locus $y_2(\phi)$.

%%%%%%%%%%%%%%%%%%%%%%%%%%%%%%%%%%%%%%%%%%%%%%%%%%%%%%%%%%%%%%%%%%%%%%%%%%%%%%%%%%

\section{Potts zeros}

From the equation (8) the locus of the Potts zeros is obtained to be
\be
Q_1(\psi)=1-a+\bigg[\sqrt{g_1}\cos\psi+i\sqrt{g_2}\bigg]^2,
\ee
where $g_1=(a-1)x$ and $g_2=g_1\sin^2\psi+x-1$.
The edge zeros of $Q_1(\psi)$ are given by
\be
Q_\pm=(a-2)(x-1)\pm2\sqrt{(1-a)x(x-1)}
\ee
from $Q_1(\pi)$ and $Q_1(0)$.
If $g_1>0$ and $g_2>0$ or $g_1<0$ and $g_2<0$,
it is easily verified that
\be
|Q_1(\psi)-(1-a)|=|ax-1|.
\ee
From Eq.~(81) we see that the zeros of $Q_1(\psi)$ lie on a circle
\be
Q_1(\psi)=Q_c+S e^{i\theta(\psi)}
\ee
in the complex $Q$ plane with center
\be
Q_c=1-a
\ee
and radius
\be
r=|S|,
\ee
where
\be
S=ax-1.
\ee
The argument $\theta(\psi)$ is given by
\be
\cos\theta(\psi)=2{(a-1)x\over ax-1}\cos^2\psi-1.
\ee
The one point of the circle $Q_1(\psi)$,
\be
Q_1\bigg(\psi={\pi\over2}\ {\rm or}\ \theta=\pi\bigg)=2-a(x+1),
\ee
always lies on the real axis.
However, if $g_1>0$ and $g_2<0$ or $g_1<0$ and $g_2>0$,
the zeros of $Q_1(\psi)$ lie on the real axis.

On the other hand, the equation (12) gives the second locus of the Potts zeros
\be
Q_2(\phi)={(a-1)(x+e^{i2\phi})-(ax+a-2)e^{i\phi}\over e^{i\phi}-x}.
\ee
The two points of $Q_2(\phi)$,
\be
Q_2(0)=1
\ee
and
\be
Q_2(\pi)={x+3\over x+1}-2a,
\ee
can be lie on the real axis.
From $|\lambda_\pm|=|\lambda_0|$ we also obtain
\be
Q_\ast={(a-1)(1-x)\over x}.
\ee

%%%%%%%%%%%%%%%%%%%%%%%%%%%%%%%%%%%%%%%%%%%%%%%%%%%%%%%%%%%%%%%%%%%%%%%%%%%%%%%%

\subsection{Three special cases: $x=1$, $x=0$, and $a=1$}

In the absence of an external field ($H_q=0$ or $x=1$),
the locus $Q_1(\psi)$ becomes
\be
Q_1(\psi)=1-a+(a-1)e^{i\theta},
\ee
where the argument $\theta(\psi)$ is simply given by
\be
\theta(\psi)=2\psi.
\ee
At the same time, the locus $Q_2(\phi)$ reduces to
\be
Q_2(\phi)=(a-1)(e^{i\phi}-1),
\ee
which is the identical circle to the locus $Q_1(\psi)$.
The equation (93) means that the Potts zeros are uniformly distributed
on this circle.
In particular, the Potts zeros lie on the unit circle at $a=0$
\cite{shrock97a,shrock97b}.
The circle $Q_1(\psi)$ or $Q_2(\phi)$ cuts the real axis at two points
$Q=0$ and $2(1-a)$.
For ferromagnetic interaction $J>0$ the physical interval is
\be
1\le a\le\infty\ \ (\infty\ge T\ge0),
\ee
whereas for antiferromagnetic interaction $J<0$ the physical interval
\be
0\le a\le1\ \ (0\le T\le\infty).
\ee
It should be noted that the point $2(1-a)$ lies on the positive real axis
for the Potts antiferromagnet.
For ${1\over2}\le a\le1$ the point $2(1-a)$ locates in $0\le2(1-a)\le1$,
while for $0\le a<{1\over2}$ it does in $1<2(1-a)\le2$.
The locus of the Potts zeros consists of the circle and the isolated point
at $Q=1$ for $0\le a<{1\over2}$.

In the limit $H_q\to-\infty$ ($x\to0$), the partition function
of the $Q$-state Potts model is again given by Eq.~(60),
and the symmetry of the Hamiltonian is that of the
$(Q-1)$-state Potts model in zero external field.
At $x=0$, the locus $Q_2(\phi)$ (Eq.~(88)) of the $Q$-state Potts model
reduces to
\be
Q_2(\phi)=2-a+(a-1)e^{i\phi},
\ee
which is a circle with center $Q_c=2-a$ and radius $r=|a-1|$.
The equation (97) is also obtained by replacing $Q$ in Eq.~(94) with $Q-1$.
The circle $Q_2(\phi)$ cuts the real axis at two points $Q=1$ and $3-2a$.
For ${1\over2}\le a\le1$ the point $3-2a$ locates in $1\le3-2a\le2$,
while for $0\le a<{1\over2}$ it does in $2<3-2a\le3$.
The locus of the Potts zeros consists of the circle $Q_2(\phi)$
and the isolated point at $Q=2$ for $0\le a<{1\over2}$.
The isolated point $Q=2$ is also obtained by replacing $Q$ in the isolated
point $Q=1$ for $x=1$ with $Q-1$.

At $a=1$ ($T=\infty$), the eigenvalue $\lambda_0=a-1$ disappears.
Therefore, the Potts zeros are completely determined by the locus $Q_1(\psi)$.
All the Potts zeros lie on the point
\be
Q_1(\psi)=1-x
\ee
for $a=1$.

%%%%%%%%%%%%%%%%%%%%%%%%%%%%%%%%%%%%%%%%%%%%%%%%%%%%%%%%%%%%%%%%%%%%%%%%%%%%%%%%

\subsection{$a>1$ and $x>1$}

In this case $g_1$ and $g_2$ are always positive and
the Potts zeros of the ferromagnetic model (ferromagnetic Potts zeros),
$Q_1(\psi)$, lie on a circle where the eigenvalues satisfy
\be
{|\lambda_\pm|\over|\lambda_0|}=\sqrt{(ax-1)x\over a-1}\ \ >1,
\ee
which implies that the locus $Q_2(\phi)$ does not appear.
The circle $Q_1(\psi)$ always cuts the real axis
at the point $Q_1({\pi\over2})$ $(<0)$.
Figure 8 shows the loci of the Potts zeros in the complex $Q$ plane for $a=3$
which are centered at
\be
Q_c=-2.
\ee
The radius of the circle $Q_1(\psi)$ is
\be
r=2\ (x=1),\ 5\ (x=2),\ 8\ (x=3)
\ee
for $a=3$.
For $x=1$ two edge zeros $Q_\pm$ cut the real axis at the origin,
whereas they move away from the origin as $x$ increases.
The Potts edge zeros are determined by the edge angle
\be
\theta_0=\theta(\psi=0)=\cos^{-1}\Bigg[{(a-2)x+1\over ax-1}\Bigg],
\ee
and for $a=3$ the edge zeros are located at
\be
Q_\pm=0\ (x=1),
\ee
\be
Q_\pm=1\pm4i\ (x=2),
\ee
and
\be
Q_\pm=2\pm4\sqrt{3}i\ (x=3).
\ee

On the circle $Q_1(\psi)$, the density of zeros $g(\theta(\psi))$
is given by
\be
g(\theta)={|{\sin{\theta\over2}|}\over2\pi
\sqrt{\sin^2{\theta\over2}-\sin^2{\theta_0\over2}}}
\ee
for $|\theta|\ge\theta_0$, and
\be
g(\theta)=0
\ee
for $|\theta|<\theta_0$.
For $x=1$, $\theta_0=0$, and the Potts zeros are distributed on a circle
with the uniform density of zeros
\be
g(\theta)={1\over2\pi}.
\ee
However, for $x>1$ the density of zeros at the Potts edge zeros $Q_\pm$
diverges, that is,
\be
g(\theta\sim\theta_0)\sim{1\over\sqrt{Q(\theta)-Q_\pm(\theta_0)}}.
\ee
In this case, the Potts edge zero can be called the Potts edge
{\it singularity} with the edge critical exponent
$\mu_e$ ($=-{1\over2}$), and the density of zeros is expressed as
\be
g(\theta\sim\theta_0)\sim(Q-Q_\pm)^{\mu_e}.
\ee

%%%%%%%%%%%%%%%%%%%%%%%%%%%%%%%%%%%%%%%%%%%%%%%%%%%%%%%%%%%%%%%%%%%%%%%%%%%%%%%%%

\subsection{$a>1$ and $x<1$}

For $\bar{a}_1<a<\bar{a}_2$, where
\be
\bar{a}_1={3x+1\over(x+1)^2}
\ee
and
\be
\bar{a}_2={1\over1-x},
\ee
the locus of the ferromagnetic Potts zeros consists of
the loop $Q_2(\phi)$ ($\phi_\ast\le\phi\le2\pi-\phi_\ast$) and
the line $Q_1(\psi)$ ($\psi_\ast\le\psi\le\pi$) on the positive real axis
between $Q_\ast=Q_1(\psi_\ast)$ and $Q_+=Q_1(0)$ ($0<Q_\ast<Q_+$).
The loop $Q_2$ meets with the line $Q_1$ at the point
\be
Q_\ast=Q_2(\phi_\ast)=Q_2(2\pi-\phi_\ast),
\ee
where $|\lambda_+|=|\lambda_-|=|\lambda_0|$.
The loop $Q_2(\phi)$ cuts the real axis at two points $Q_2(\pi)$ and
$Q_\ast$ ($>Q_2(\pi)$). The sign of $Q_2(\pi)$
is positive for $\bar{a}_1<a<\bar{a}_3$, where
\be
\bar{a}_3={x+3\over2(x+1)},
\ee
and negative for $\bar{a}_3<a<\bar{a}_2$.
Figure 9 shows the locus of the Potts zeros for $a={3\over2}$ and $x={1\over2}$
from which we obtain
\be
Q_2(\pi)=-{2\over3},\ Q_\ast={1\over2},\ Q_+={1\over4}+{1\over\sqrt{2}}=0.957.
\ee

At $a=\bar{a}_1$, $\phi_\ast=\pi$,
the other edge zero $Q_-$ appears, and the loop $Q_2(\phi)$ shrinks
to the point
\be
Q_-=Q_\ast=Q_2(\pi)=\bigg({x-1\over x+1}\bigg)^2.
\ee
For $a<\bar{a}_1$, the only locus is the line $Q_1(\psi)$
on the positive real axis between $Q_-$ and $Q_+$ ($0<Q_-<Q_+$).
On the other hand, at $a=\bar{a}_2$, $\phi_\ast=0$, and the line $Q_1(\psi)$
shrinks to the point
\be
Q_+=Q_\ast=Q_2(0)=1.
\ee
For $a>\bar{a}_2$, all the Potts zeros
lie on the loop $Q_2(\phi)$ which again cuts the real axis at two points
$Q_2(\pi)$ and $Q_2(0)$ ($Q_2(\pi)<Q_2(0)=1$).
The sign of $Q_2(\pi)$ is positive for $\bar{a}_2<a<\bar{a}_3$
and negative for $a>\bar{a}_3$.

%%%%%%%%%%%%%%%%%%%%%%%%%%%%%%%%%%%%%%%%%%%%%%%%%%%%%%%%%%%%%%%%%%%%%%%%%%%%%

\subsection{$a<1$ and $x>1$}

Because $g_1$ is always negative, the zeros of $Q_1(\psi)$ lie on a circle
if $g_2<0$ ($\psi_0<\psi<\pi-\psi_0$) and on the real axis
if $g_2>0$ ($0\le\psi<\psi_0$ or $\pi-\psi_0<\psi\le\pi$).
For $0\le a<\bar{a}_4$, where
\be
\bar{a}_4={1\over x+1},
\ee
the locus of the Potts zeros of the antiferromagnetic model (antiferromagnetic
Potts zeros) consists of the line $Q_1(\psi)$
($0\le\psi\le\psi_0$ and $\pi-\psi_0\le\psi\le\psi_\ast$) on the
real axis between $Q_-=Q_1(0)$ ($<0$) and $Q_\ast=Q_1(\psi_\ast)$ ($>0$),
the circle $Q_1(\psi)$ ($\psi_0\le\psi\le\pi-\psi_0$),
and the loop $Q_2(\phi)$
($-\phi_\ast\le\phi\le\phi_\ast$),
inside the circle $Q_1$. The loop $Q_2$ meets with the line $Q_1$ at
the point
\be
Q_\ast=Q_2(\phi_\ast)=Q_2(-\phi_\ast),
\ee
where $|\lambda_+|=|\lambda_-|=|\lambda_0|$.
The circle $Q_1(\psi)$ cuts the real axis at two points $Q_1^0$ ($\ge0$) and
$Q_1\big({\pi\over2}\big)$ ($1<Q_1\big({\pi\over2}\big)\le2$),
where the points $Q_1^0$ and $Q_1\big({\pi\over2}\big)$ are given by
\be
Q_1^0=Q_1(\psi_0)=Q_1(\pi-\psi_0)=Q_c+S=a(x-1)
\ee
and
\be
Q_1\bigg({\pi\over2}\bigg)=Q_c-S=2-a(x+1),
\ee
respectively. Similarly, the loop also cuts the real axis at two points
$Q_\ast$ and $Q_2(0)$ ($0<Q_\ast<Q_2(0)=1$).
For example, figure 10 shows the locus for $a={1\over10}$ and $x=2$.
In this case we obtain
\be
Q_-=-{19\over10}-{6\over\sqrt{5}}=-4.583,\
Q_1^0={1\over10},\ Q_\ast={9\over20},\ Q_1\bigg({\pi\over2}\bigg)={17\over10},
\ee
and the center $Q_c$ and the radius $r$ for the circle
\be
Q_c={9\over10},\ r={4\over5}.
\ee

At $a=\bar{a}_4$, two points $Q_\ast$ and $Q_1^0$ on the real axis meet at
\be
Q_\ast=Q_1^0={x-1\over x+1},
\ee
other three points $Q_1({\pi\over2})$, $Q_2(0)$ and $Q_2(\pi)$
on the real axis also meet at
\be
Q_1\bigg({\pi\over2}\bigg)=Q_2(0)=Q_2(\pi)=1,
\ee
and the circle $Q_1(\psi)$ and the loop $Q_2(\phi)$
become the identical locus as a circle with center
\be
Q_c={x\over x+1}
\ee
and radius
\be
r={1\over x+1}.
\ee
On this circle three eigenvalues have the same magnitude
\be
|\lambda_+|=|\lambda_-|=|\lambda_0|={x\over x+1}.
\ee
Therefore, the locus consists of the line $Q_1(\psi)$ on the
real axis between $Q_-$ and $Q_\ast$ and the circle.
In the region $\bar{a}_4<a<\bar{a}_1$,
the circle $Q_1(\psi)$ disappears, and
the line $Q_1(\psi)$ ($0\le\psi\le\psi_\ast$) on the real axis
between $Q_-$ ($<0$) and $Q_\ast$ ($>0$) again meets with
the loop $Q_2(\phi)$ ($\phi_\ast\le\phi\le2\pi-\phi_\ast$) at the point $Q_\ast$.
The loop cuts the real axis at two points $Q_\ast$ and $Q_2(\pi)$
($0<Q_\ast<Q_2(\pi)<1$).

At $a=\bar{a}_1$, the other edge zero $Q_+$ appears,
and the loop $Q_2(\pi)$ shrinks to the point
\be
Q_+=Q_\ast=Q_2(\pi)=\bigg({x-1\over x+1}\bigg)^2.
\ee
For $a>\bar{a}_1$,
the loop disappears, and the only locus is the line $Q_1(\psi)$ on the real axis
between $Q_-$ ($<0$) and $Q_+$ ($>Q_-$).
The sign of $Q_+$ is positive (negative) if $a<\bar{a}_5$ ($a>\bar{a}_5$), where
\be
\bar{a}_5=-{2(1-\sqrt{x})\over x+1}.
\ee
In the limit $x\to\infty$, the boundary curves $a=\bar{a}_1$
and $a=\bar{a}_4$ approach the line $a=0$, and the Potts zeros
lie on the line $Q_1(\psi)$ which approaches the point $-\infty$.

%%%%%%%%%%%%%%%%%%%%%%%%%%%%%%%%%%%%%%%%%%%%%%%%%%%%%%%%%%%%%%%%%%%%%%%%%%%%%%%%%

\subsection{$a<1$ and $x<1$}

For $a<1$ and $x<1$, $g_1$ and $g_2$ are always negative, and
the antiferromagnetic Potts zeros of the locus $Q_1(\psi)$ lie on a circle.
At $a=\bar{a}_4$, the loci $Q_1(\psi)$ and $Q_2(\phi)$
become identical to be a circle on which
three eigenvalues again have the same magnitude
\be
|\lambda_+|=|\lambda_-|=|\lambda_0|={x\over x+1}.
\ee
This circle cuts the real axis only at the point
\be
Q_1\bigg({\pi\over2}\bigg)=Q_2(0)=Q_2(\pi)=1.
\ee
For $a>\bar{a}_4$, all the Potts zeros lie on the circle $Q_1(\psi)$
($0\le\psi\le\pi$) which cuts the real axis at the point $Q_1({\pi\over2})$
($0<Q_1({\pi\over2})<1$).
Figure 11 shows the locus of the Potts zeros for $a={3\over5}$ and $x={9\over10}$
which has the center
\be
Q_c={2\over5},
\ee
the radius
\be
r={23\over50},
\ee
the edge zeros
\be
Q_\pm={7\over50}\pm i{3\over5}\sqrt{2\over5}=0.140\pm0.379i,
\ee
and the point on the positive real axis
\be
Q_1\bigg({\pi\over2}\bigg)={43\over50}.
\ee

On the other hand,
for $0\le a<\bar{a}_4$, the locations of Potts zeros are completely
determined by the loop $Q_2(\phi)$ ($0\le\phi\le2\pi$)
which cuts the real axis at two points $Q_2(0)$ ($=1$)
and $Q_2(\pi)$ ($1<Q_2(\pi)<3$).
The locus $Q_2(\phi)$ has a crescent-like shape near $a=\bar{a}_4$,
but an egg-like shape far away from $a=\bar{a}_4$.
Figures 12 and 13 show the loci of the Potts zeros
for $a={1\over2}$ and $x={1\over2}$
and for $a={1\over10}$ and $x={1\over10}$, respectively.
Figure 12 shows a crescent-like shape with
\be
Q_2(\pi)={4\over3},
\ee
while figure 13 has an egg-like shape and
\be
Q_2(\pi)={144\over55}.
\ee
For $\bar{a}_6\le a<\bar{a}_4$, where
\be
\bar{a}_6={1-x\over2(x+1)},
\ee
the point $Q_2(\pi)$ locates in
$1<Q_2(\pi)\le2$, while for $0\le a<\bar{a}_6$ it does in $2<Q_2(\pi)<3$.
The locus of the Potts zeros consists of the loop $Q_2(\phi)$ and
the isolated point at $Q=2$ for $0\le a<\bar{a}_6$.

%%%%%%%%%%%%%%%%%%%%%%%%%%%%%%%%%%%%%%%%%%%%%%%%%%%%%%%%%%%%%%%%%%%%%%%%%%

\section{conclusion}

We have investigated the properties of the Fisher zeros
in the complex temperature plane and the Potts zeros in the complex $Q$ plane
of the $Q$-state Potts model in an arbitrary nonzero external magnetic field,
finding two master equations to determine the partition
function zeros from the exact partition function of the one-dimensional model.
The distribution of the Fisher zeros or the Potts zeros is determined
by the interplay between these master equations.
We have discussed the Fisher zeros of the Potts model
in an external magnetic field for any real value of $Q\ge0$.
We have also studied the Potts zeros in the complex $Q$ plane
of the ferromagnetic and antiferromagnetic Potts models for nonzero magnetic field.

Some circle theorems have been found for both the Fisher zeros and the Potts zeros.
All Fisher zeros lie on a circle for $Q>1$ and $x\ge1$ except $Q=2$ (Ising model)
whose zeros lie on the imaginary axis.
All Fisher zeros also lie on a circle for any value of $Q$ when $x=1$
(except $Q=0$, 1 and 2) or $x=0$ (except $Q=1$, 2 and 3).
All Potts zeros of the ferromagnetic model lie on a circle for $x\ge1$.
When $x=1$ or $x=0$, all Potts zeros lie on a circle for both
the ferromagnetic and antiferromagnetic models.
All Potts zeros of the antiferromagnetic model with $x<1$ also lie on a circle
for $(x+1)^{-1}\le a<1$.
It has been found that a part of the Fisher zeros or the Potts zeros lie on
a circle for the specific ranges of $x$.

We have shown that some Fisher or Potts zeros can cut the positive real axis.
Furthermore, the Fisher zeros or the Potts zeros lie on the positive real axis
for the specific ranges of $x$.
We have also calculated and discussed the density of zeros.
The density of zeros at the Fisher edge singularity diverges, and the edge
critical exponents at the singularity satisfy the Rushbrooke scaling law.
We have found the Potts edge singularity in the complex $Q$ plane
which is similar to the Fisher edge singularity in the complex temperature plane.

%%%%%%%%%%%%%%%%%%%%%%%%%%%%%%%%%%%%%%%%%%%%%%%%%%%%%%%%%%%%%%%%%%%%%%%%%%

\begin{acknowledgments}
The author thanks Prof. R. J. Creswick for valuable discussions.
\end{acknowledgments}

%%%%%%%%%%%%%%%%%%%%%%%%%%%%%%%%%%%%%%%%%%%%%%%%%%%%%%%%%%%%%%%%%%%%%%%%%%%%%%

%%%%%%%%%%%%%%%%%%%%%%%%%%%%%%%%%%%%%%%%%%%%%%%%%%%%%%%%%%%%%%%%%%%%%%%%%%
\newpage

\begin{figure}
\includegraphics{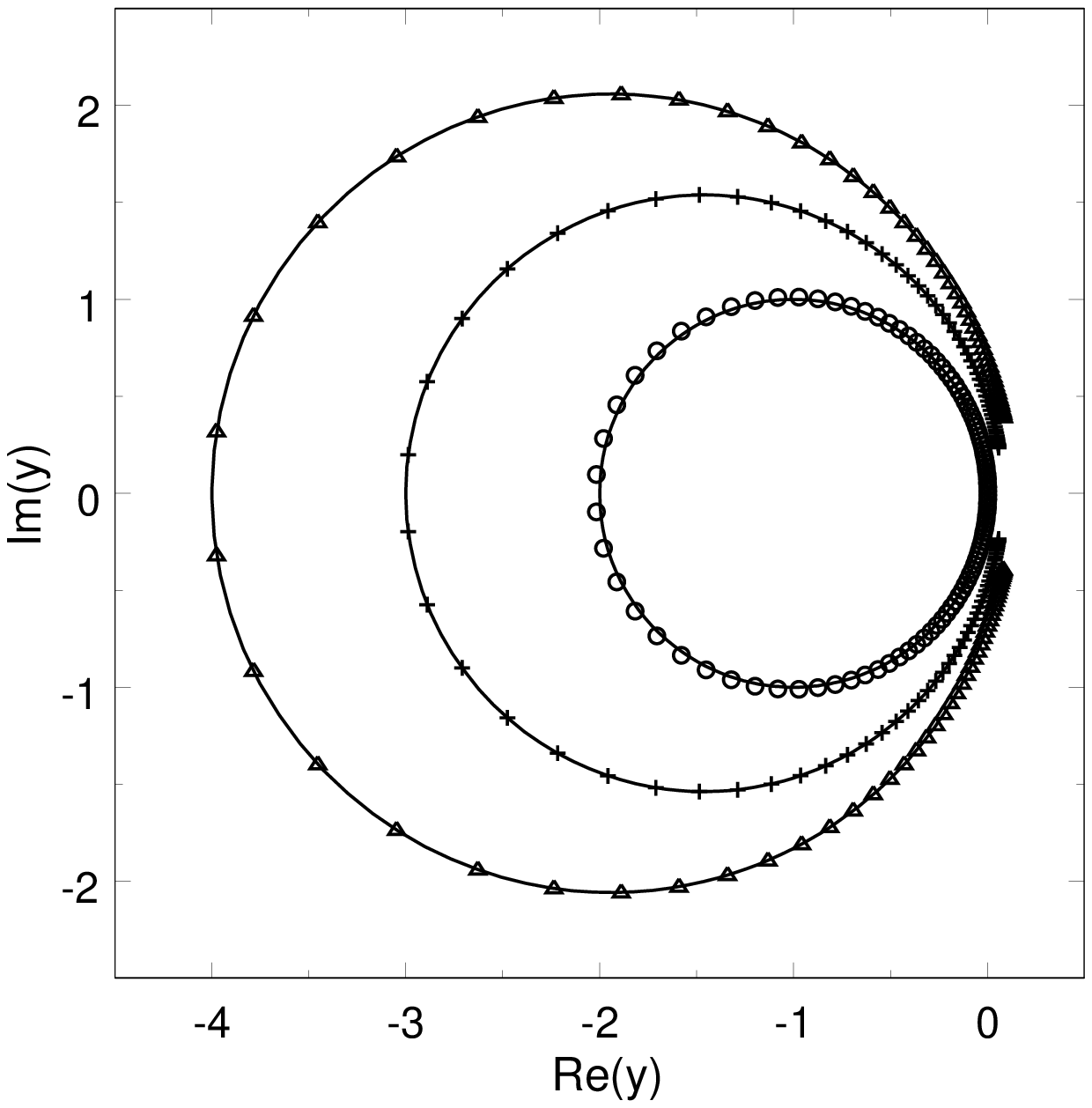}
\caption{The locus of the Fisher zeros in the complex $y=e^{-\beta J}$ plane
of the three-state Potts model for $x=e^{\beta H_q}=1$, 2 and 3.
For comparison, the zeros of a finite-size system ($N=100$) are also shown
(open circles for $x=1$, pluses for $x=2$,
and open triangles for $x=3$, respectively).}
\end{figure}

\begin{figure}
\includegraphics{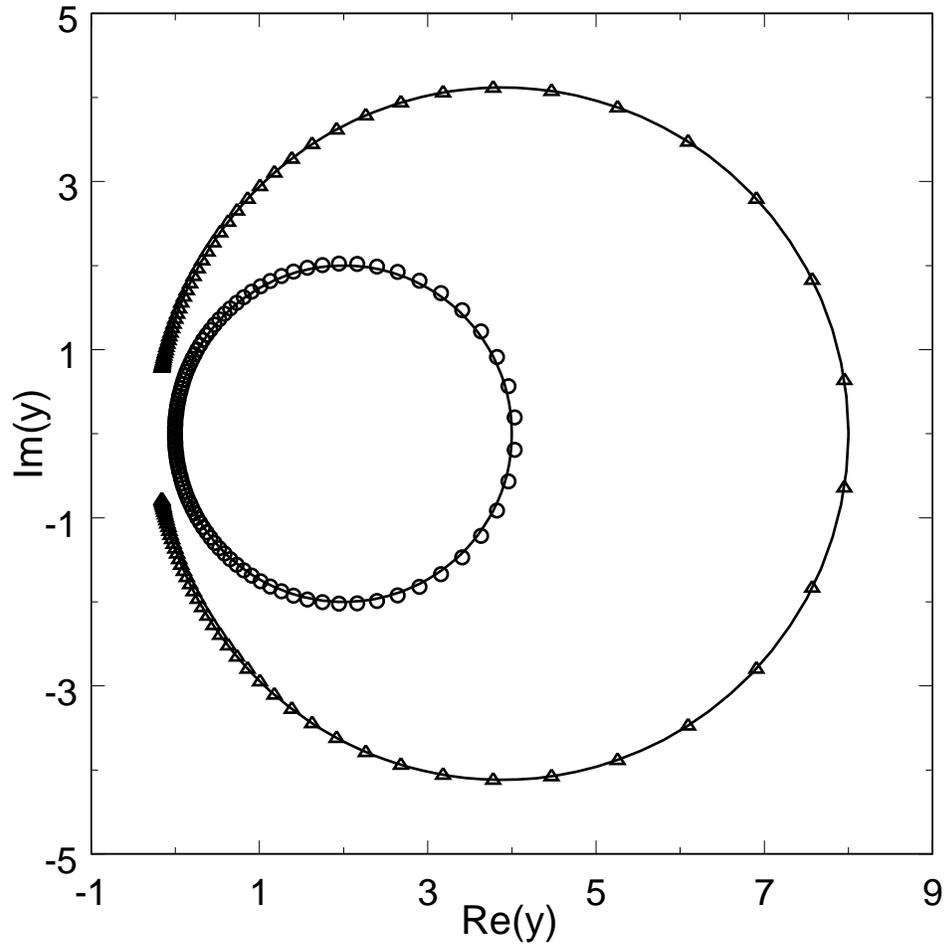}
\caption{The locus of the Fisher zeros of the $Q={3\over2}$ state Potts model
for $x=1$ (open circles) and $x=3$ (open triangles).}
\end{figure}

\begin{figure}
\includegraphics{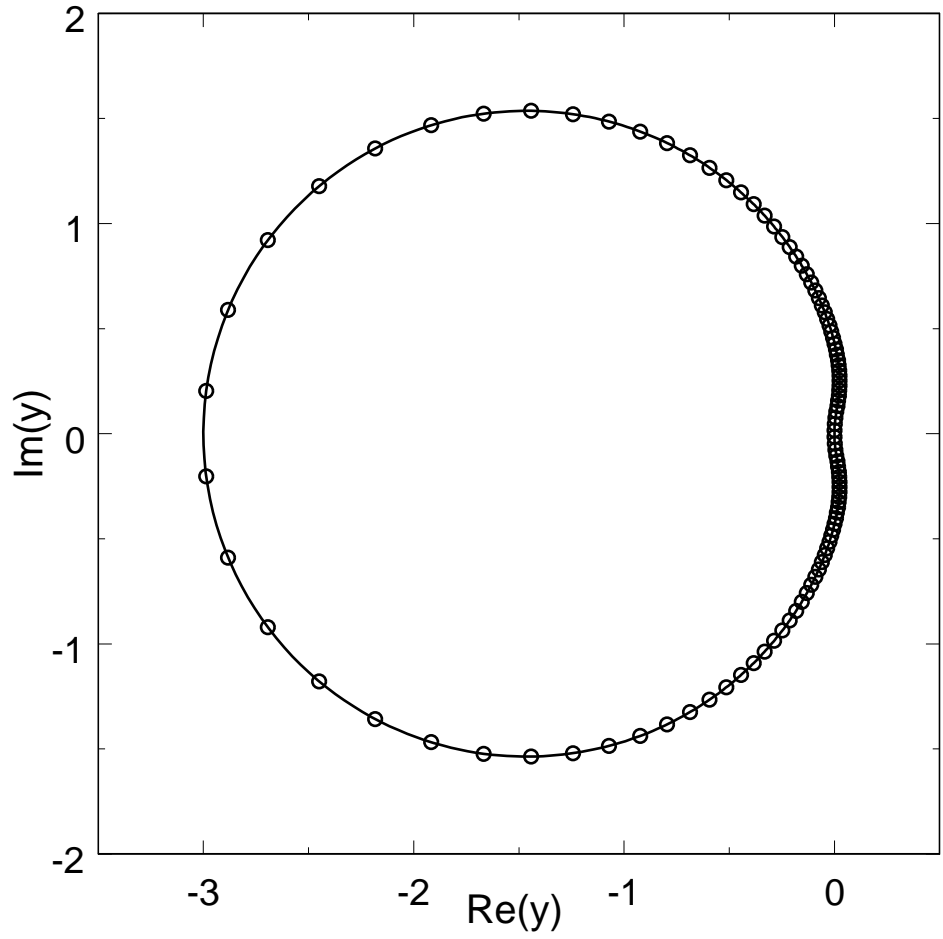}
\caption{The locus of the Fisher zeros of the three-state Potts model
for $x={1\over2}$.}
\end{figure}

\begin{figure}
\includegraphics{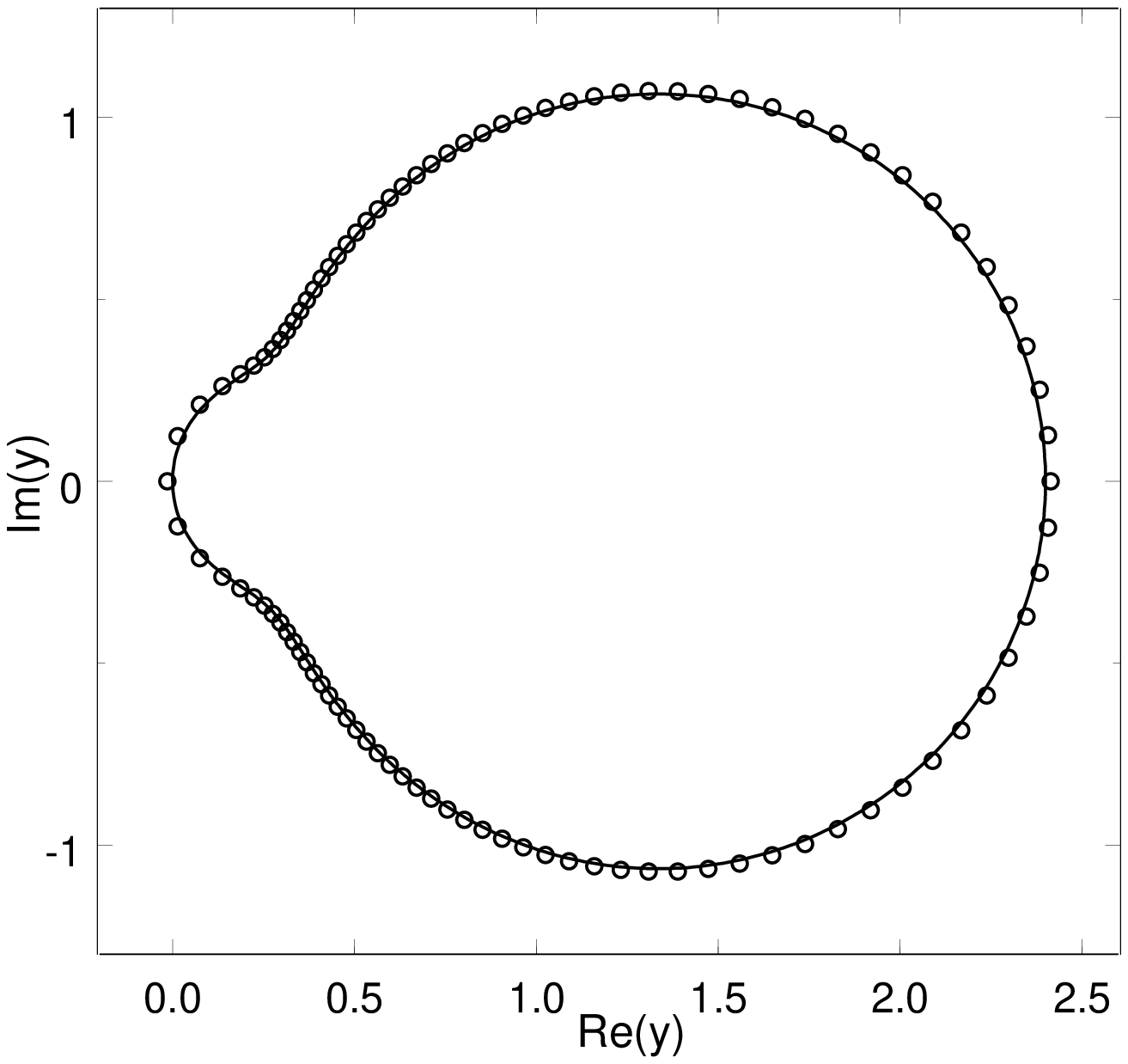}
\caption{The locus of the Fisher zeros of the $Q={3\over2}$ state Potts model
for $x={1\over2}$.}
\end{figure}

\begin{figure}
\includegraphics{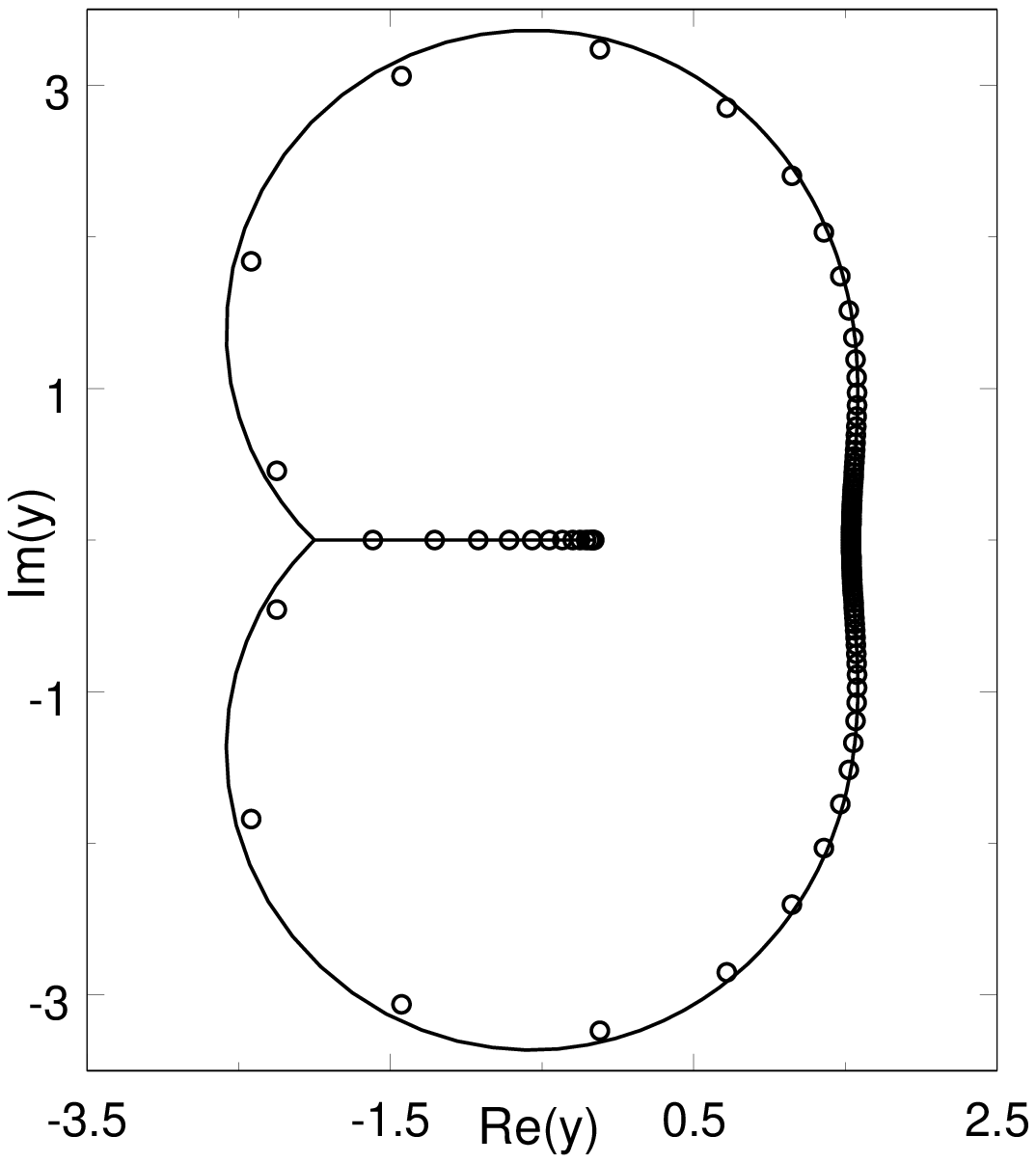}
\caption{The locus of the Fisher zeros of the $Q={1\over2}$ state Potts model
for $x={3\over2}$.}
\end{figure}

\begin{figure}
\includegraphics{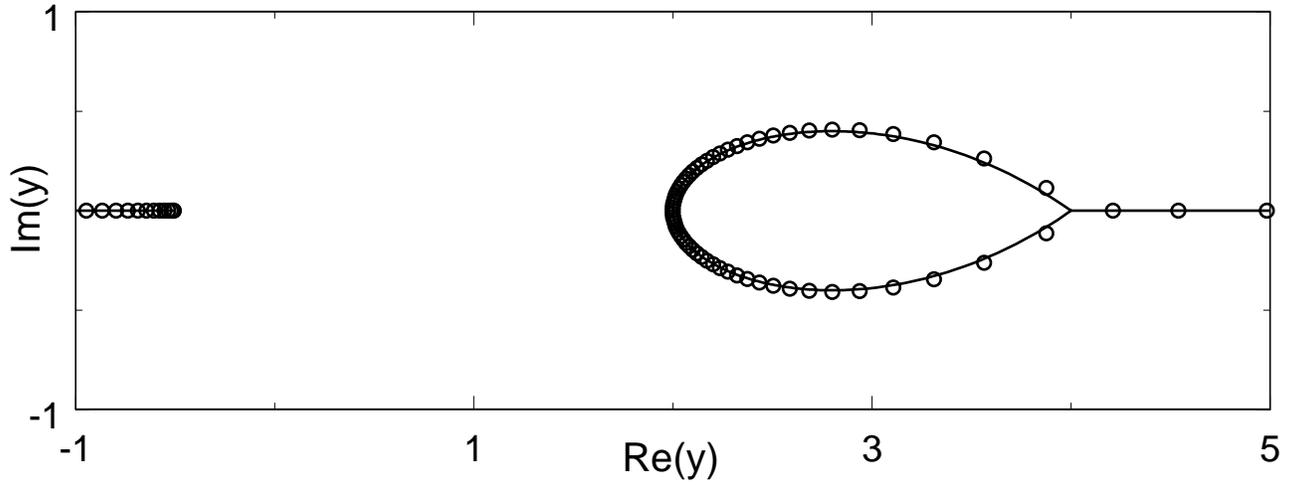}
\caption{The locus of the Fisher zeros of the $Q={1\over2}$ state Potts model
for $x=3$. The zeros on the real axis lie between $-\infty$ and $y_+=-0.506$
and between $y_\ast=4$ and $\infty$. Most of them are omitted in the figure.}
\end{figure}

\begin{figure}
\includegraphics{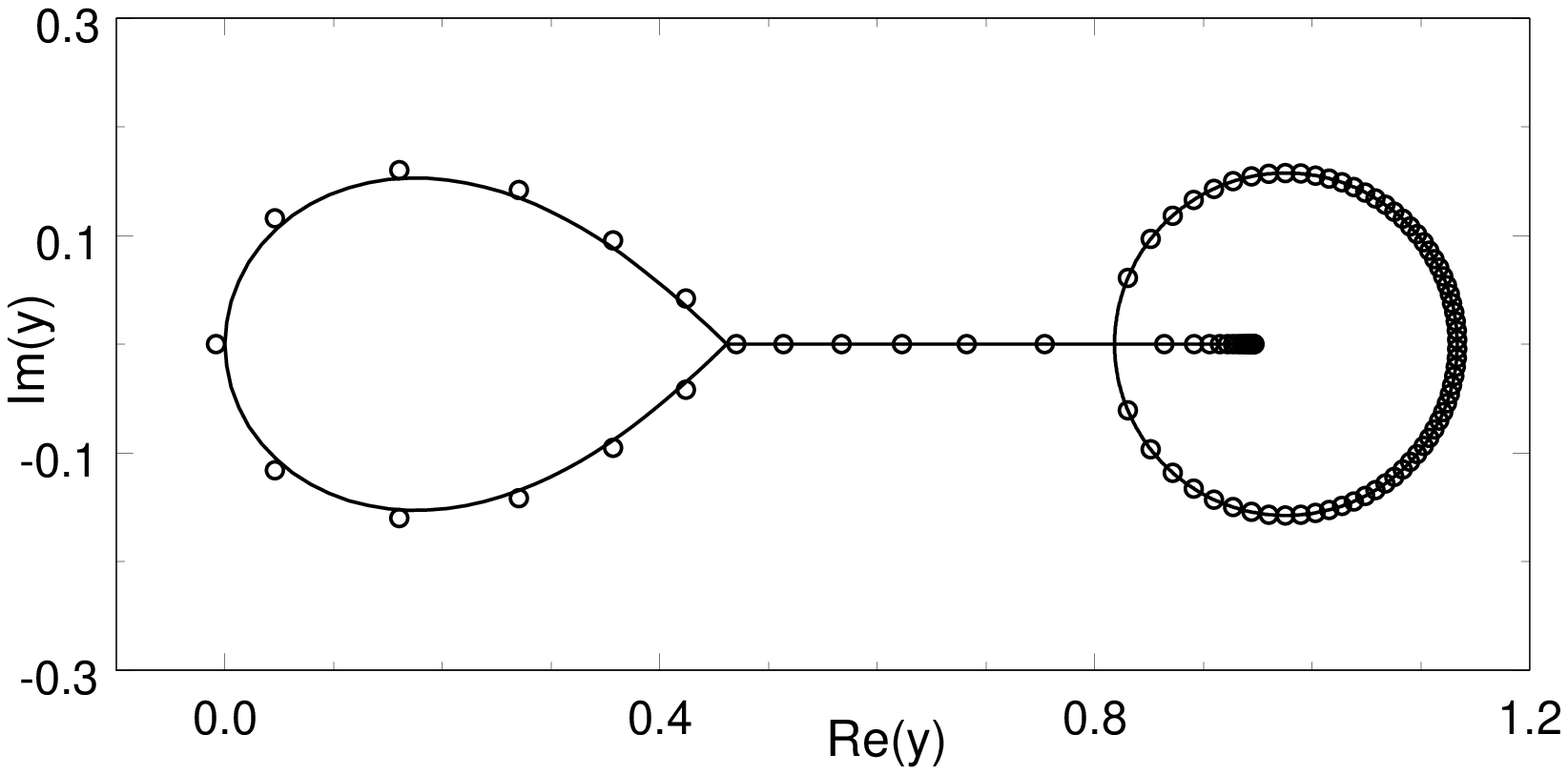}
\caption{The locus of the Fisher zeros of the $Q={1\over2}$ state Potts model
for $x={7\over10}$.}
\end{figure}

\begin{figure}
\includegraphics{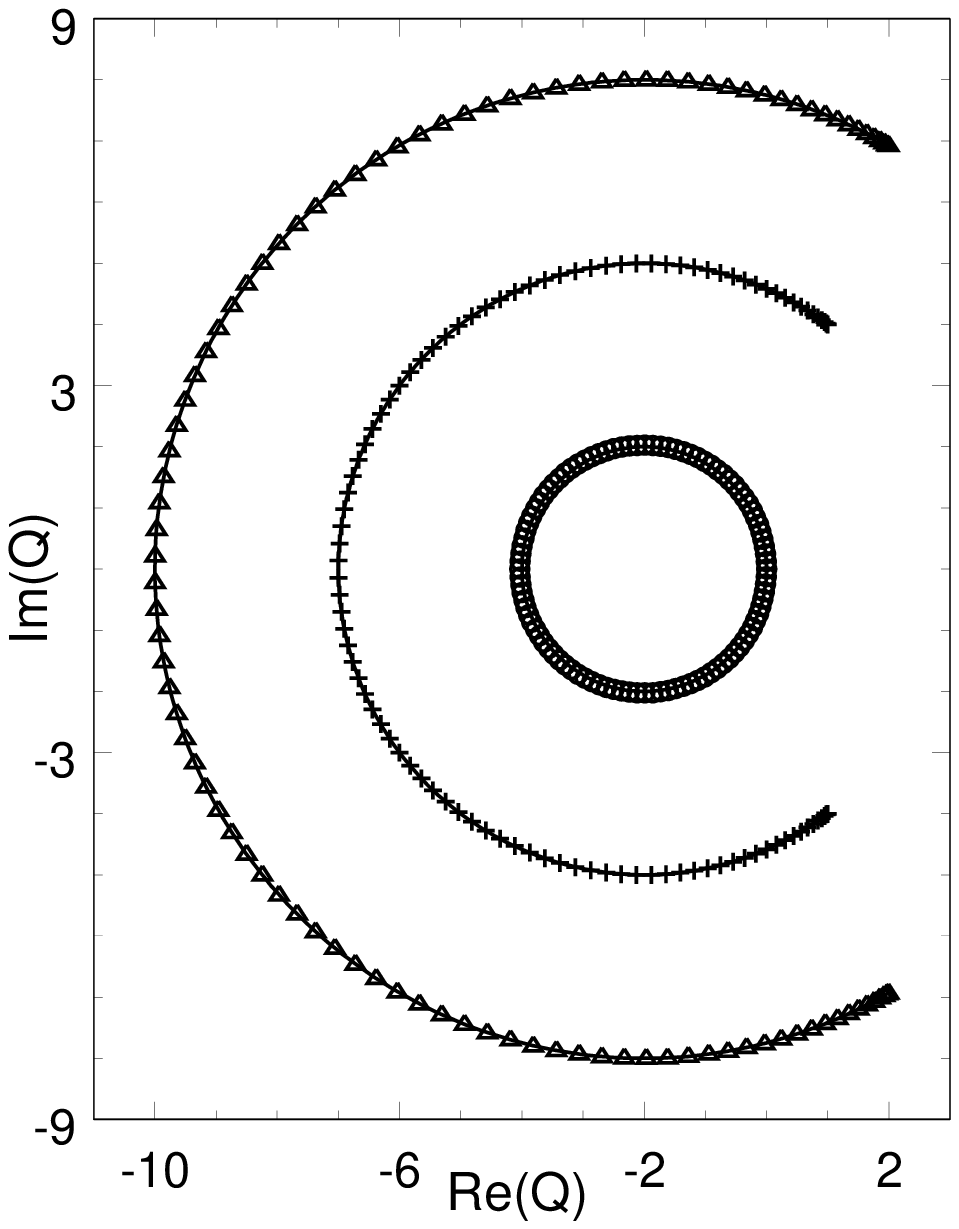}
\caption{The locus of the Potts zeros in the complex $Q$ plane for $a=3$.
For comparison, the zeros of a finite-size system ($N=100$) are also shown
(open circles for $x=1$, pluses for $x=2$,
and open triangles for $x=3$, respectively).}
\end{figure}

\begin{figure}
\includegraphics{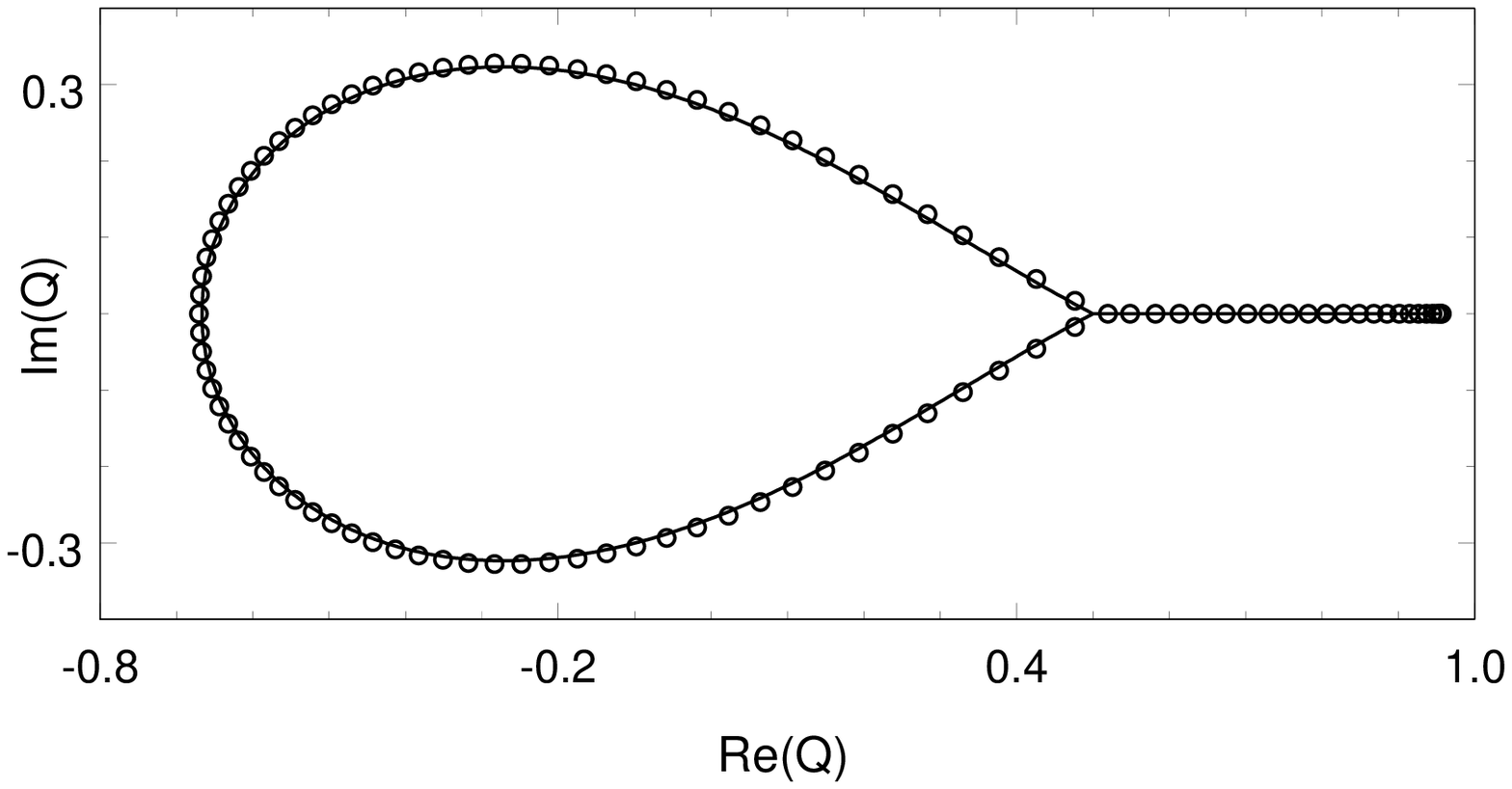}
\caption{The locus of the Potts zeros for $a={3\over2}$ and $x={1\over2}$.}
\end{figure}

\begin{figure}
\includegraphics{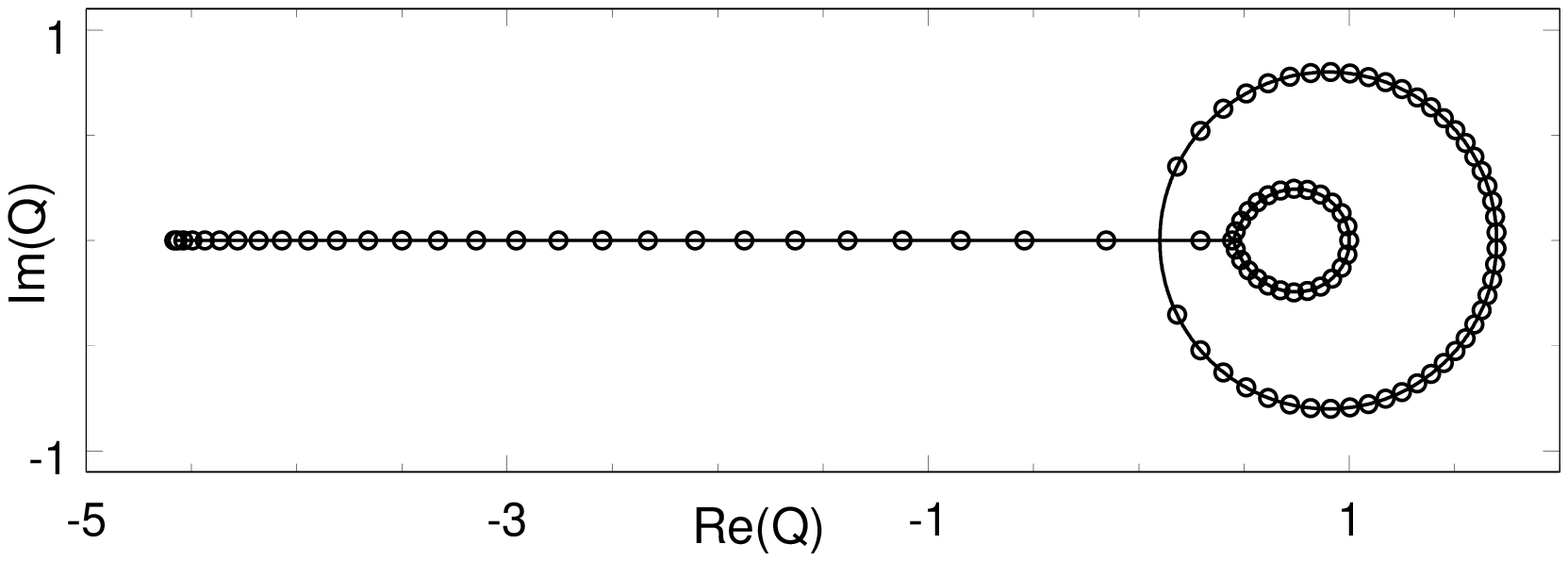}
\caption{The locus of the Potts zeros for $a={1\over10}$ and $x=2$.}
\end{figure}

\begin{figure}
\includegraphics{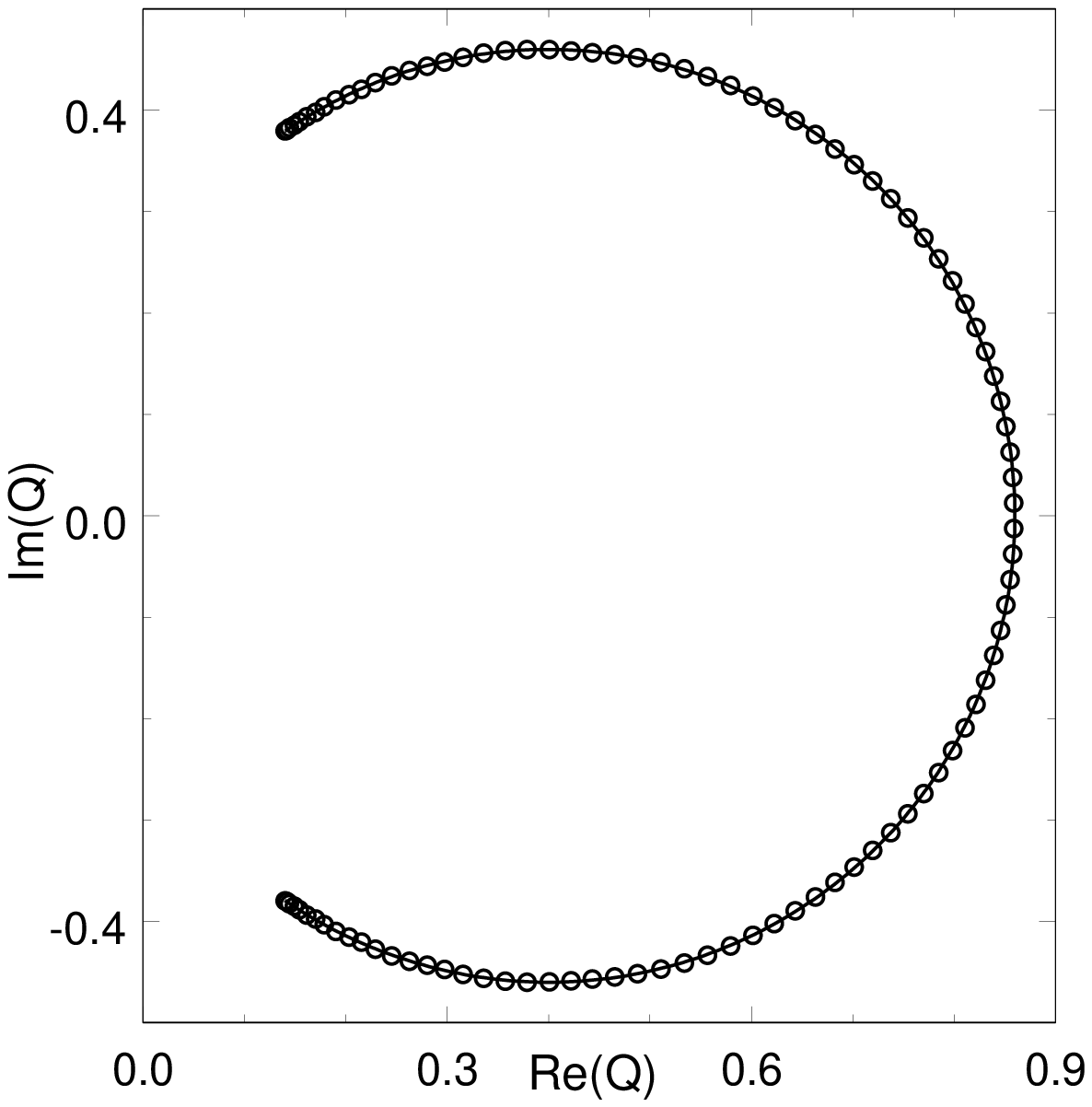}
\caption{The locus of the Potts zeros for $a={3\over5}$ and $x={9\over10}$.}
\end{figure}

\begin{figure}
\includegraphics{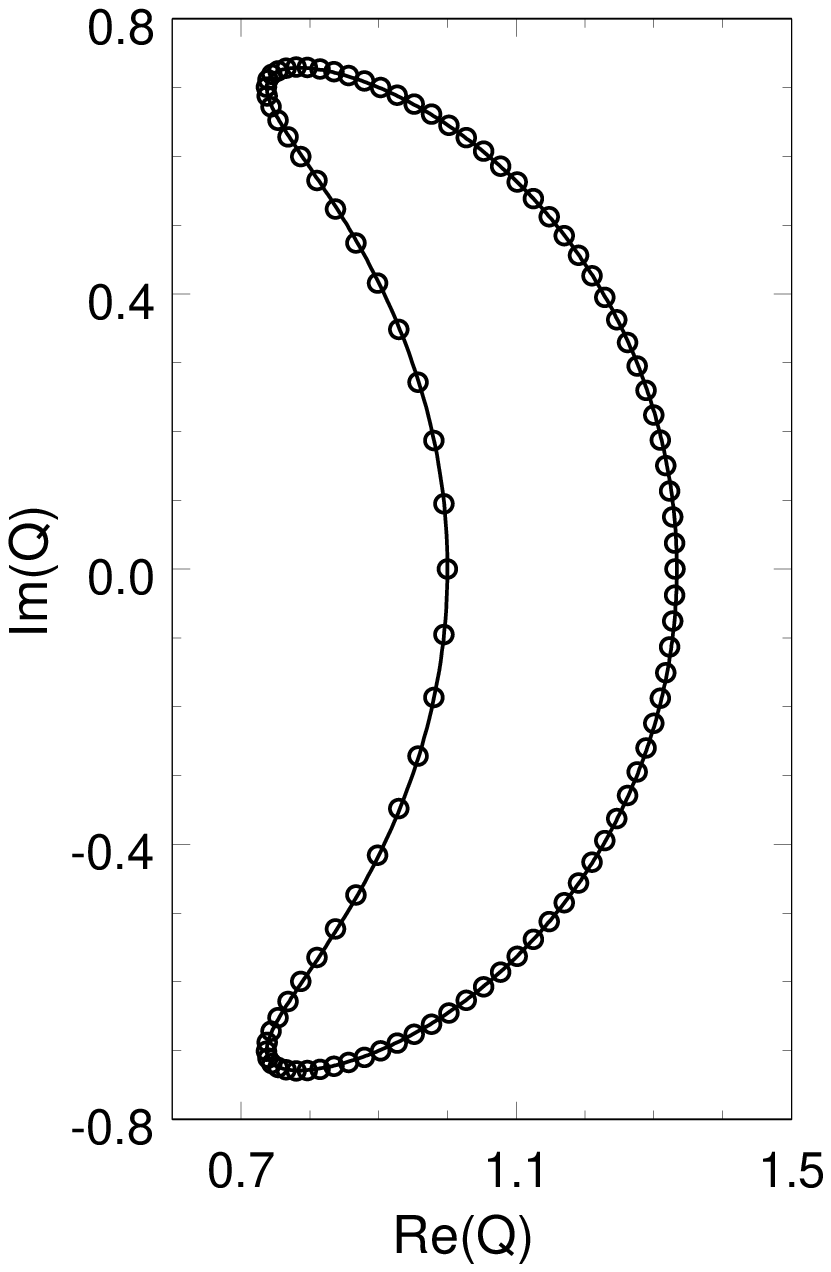}
\caption{The locus of the Potts zeros for $a={1\over2}$ and $x={1\over2}$.}
\end{figure}

\begin{figure}
\includegraphics{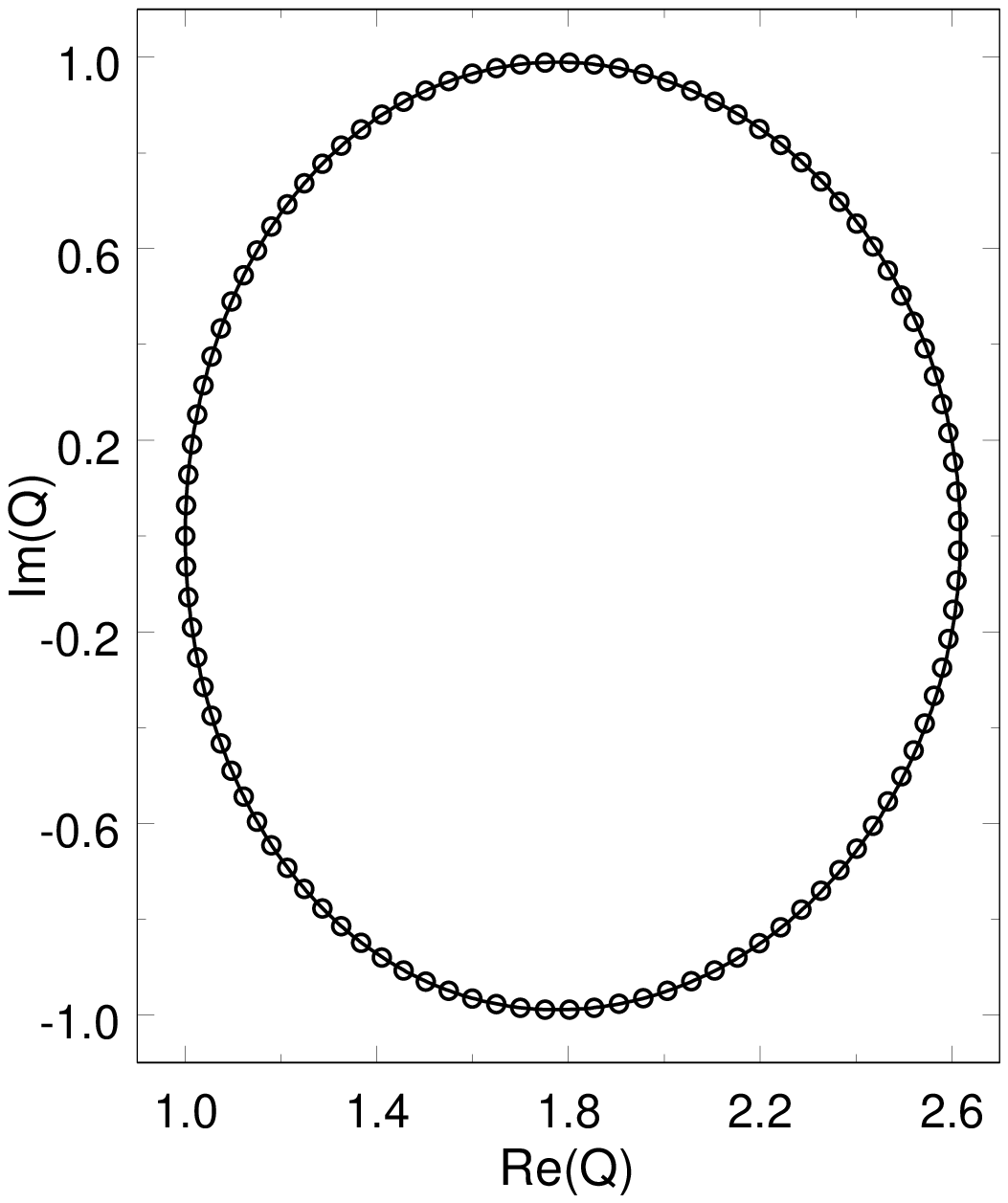}
\caption{The locus of the Potts zeros for $a={1\over10}$ and $x={1\over10}$.}
\end{figure}

%%%%%%%%%%%%%%%%%%%%%%%%%%%%%%%%%%%%%%%%%%%%%%%%%%%%%%%%%%%%%%%%%%%%%%%%%%%%%%%

\end{document}